\documentclass[fleqn,usenatbib]{mnras}

\usepackage{newtxtext,newtxmath}
\usepackage[T1]{fontenc}
\usepackage{ae,aecompl}

\input psfig.tex
\usepackage[english]{babel}
\usepackage{graphicx}
\usepackage{caption}
\usepackage{morefloats}
\usepackage{natbib}
\bibpunct{(}{)}{;}{a}{}{,}
\usepackage{array}
\usepackage{graphics}
\usepackage{latexsym}
\usepackage{amssymb}
\usepackage{amsmath}
\usepackage{fancyhdr}
\usepackage{float}
\usepackage{multirow}
\usepackage{longtable}
\usepackage{lscape}
\usepackage{morefloats}
\usepackage{slashbox}

\usepackage{graphicx}	% Including figure files
\usepackage{amsmath}	% Advanced maths commands
\usepackage{amssymb}	% Extra maths symbols

\title[Red and dead galaxies in Goods-S]{Chasing passive galaxies in the early Universe:\\a critical analysis in CANDELS GOODS-South}

\author[E. Merlin et al.]{E. Merlin$^{1}$,
A. Fontana$^{1}$, 
M. Castellano$^{1}$, 
P. Santini$^{1}$, 
M. Torelli$^{1}$, 
K. Boutsia$^{1,2}$,
\newauthor T. Wang$^{3,7,8}$, 
A. Grazian$^{1}$, 
L. Pentericci$^{1}$, 
C. Schreiber$^{4}$, 
L. Ciesla$^{3}$, 
R. McLure$^{6}$, 
\newauthor S. Derriere$^{5}$, 
J. S. Dunlop$^{6}$, 
and D. Elbaz$^{3}$
\\
\\
$^{1}$INAF - Osservatorio Astronomico di Roma, via Frascati 33, 00078 Monte Porzio Catone (RM), Italy\\
$^{2}$Carnegie Observatories, Las Campanas Observatory, Colina El Pino, Casilla 601, La Serena, Chile\\
$^{3}$Laboratoire AIM-Paris-Saclay, CEA/DSM/Irfu - CNRS - Universit\'e Paris Diderot, \\CEA-Saclay, pt courrier 131, F-91191 Gif-sur-Yvette, France\\
$^{4}$Leiden Observatory, Leiden University, 2300 RA Leiden, The Netherlands\\
$^{5}$Observatoire astronomique de Strasbourg, Universit\'e de Strasbourg, CNRS, UMR 7550, 11 rue de l'Universit\'e, F-67000 Strasbourg, France\\
$^{6}$SUPA\thanks{Scottish Universities Physics Alliance}, Institute for Astronomy, University of Edinburgh, Royal Observatory, Edinburgh, EH9 3HJ, U.K.\\
$^{7}$Institute of Astronomy, The University of Tokyo, Osawa, Mitaka, Tokyo 181-0015,
Japan\\
$^{8}$National Astronomical Observatory of Japan, Mitaka, Tokyo 181-8588, Japan
}

\date{Accepted XXX. Received YYY; in original form ZZZ}

\pubyear{2017}

\begin{document}
\label{firstpage}
\pagerange{\pageref{firstpage}--\pageref{lastpage}}
\maketitle

\begin{abstract}

We search for passive galaxies at $z$$>$3 in the GOODS-South field, using different techniques based on photometric data, and paying attention to develop methods that are sensitive to objects that have become passive shortly before the epoch of observation.
We use CANDELS HST catalogues, ultra-deep $Ks$ data and new IRAC photometry, performing spectral energy distribution fitting using models with abruptly quenched star formation histories. We then single out galaxies which are best fitted by a passively evolving model, and having only low probability (<5\%) star-forming solutions. We verify the effects of including nebular lines emission, and we consider possible solutions at different redshifts.
The number of selected sources dramatically depends on the models used in the SED fitting. Without including emission lines and with photometric redshifts fixed at the CANDELS estimate, we single out 30 candidates; the inclusion of nebular lines emission reduces the sample to 10 objects; allowing for solutions at different redshifts, only 2 galaxies survive as robust candidates. Most of the candidates are not far-infrared emitters, corroborating their association with passive galaxies. 
Our results translate into an upper limit in the number density of $\sim$0.173 arcmin$^2$ above the detection limit. However, we conclude that the selection of passive galaxies at $z$$>$3 is still subject to significant uncertainties, being sensitive to assumptions in the SED modeling adopted and to the relatively low S/N of the objects. By means of dedicated simulations, we show that \textit{JWST} will greatly enhance the accuracy, allowing for a much more robust classification.

\end{abstract}


%%%%%%%%%%%%%%%%%%%%%%%%%%%%%%%%%%%%%%%%%%%%%%%%%%

%%%%%%%%%%%%%%%%% BODY OF PAPER %%%%%%%%%%%%%%%%%%

\section{Introduction} \label{intro}

Observational data are at variance with theoretical predictions about the grand-picture of how galaxies build up their stellar mass during the early phases of their growth. While the concordance cosmological scenario postulates a bottom-up, hierarchical assembly of structures, the existence of massive, passively evolving galaxies in the early Universe is now a well-established evidence \citep[e.g.][]{Labbe2005, Mobasher2005, Rodighiero2007, Wiklind2008, Fontana2009, Marchesini2010, Muzzin2013, Stefanon2013, Nayyeri2014, Straatman2014, Grazian2015}.
Theoretical and numerical models struggle to find a way to reconcile with the observations, but the results are still not satisfactory, and fine-tuning of free parameters is often required \citep[see e.g.][]{Silk2012, Vogelsberger2014, Schaye2015, Feldmann2017}. Dedicated hydro-dynamical simulations have shown that the star formation history (SFH) of galaxies largely depends on the mass of the proto-galactic halo and, secondary, on its environment, so that large overdense regions in the early Universe can generate massive galaxies with a very short and intense burst of star formation activity, followed by quiescence ever since \citep[e.g.][]{Merlin2012}; however, it is not clear to what extent the abundance of such objects challenges large-scale theoretical predictions, and whether the extremely short formation time-scales of observed passive galaxies are compatible with the ones obtained in simulations. %Recent data on the stellar mass functions of high-redshift ($z>$4) galaxies appear to be incompatible with the halo mass functions predicted by the $\Lambda$-CDM cosmology \citep{Steinhardt2016}. As discussed by \citet{Behroozi2016}, observations at $z>10$, likely to be available in the next future with the \textit{James Webb Space Telescope} (\textit{JWST} hereafter), will be crucial to assess the reliability of the cosmological framework against the existence of very early massive objects.

What we know for sure is that these objects do exist, and they are not too rare. After the early discoveries of massive, red, passive galaxies at $z \simeq 1-2$ \citep[e.g.][]{Rosati1999, Daddi2000a, Daddi2000b, Daddi2005}, and their spectroscopic confirmation \citep[e.g.][]{Cimatti2004, vanDokkum2004, Whitaker2013}, the search for quenched sources has been extended to $z \sim 3-4$ and beyond, necessarily basing the studies on photometric samples. 

% Thanks to the unprecedented quality reached by recent deep field surveys, such as CANDELS \citep{Koekemoer2011,Grogin2011} and 3D-HST \citep{Brammer2012,Skelton2014}, it is now possible to try an extended search of such objects. 

In this paper, we use the best data available today to extend this search at the deepest attainable limits, in order to reliably detect and classify passive galaxies well above $z\simeq 3$, when the age of the Universe was less than $\sim$2 Gyr. The very definition of ``passive'' galaxy is clearly somewhat ambiguous. Very low levels of star--formation rate are hard to measure in cosmologically distant galaxies, and it is therefore difficult to establish whether a given object has a star formation rate (SFR) exactly equal to zero. Also, when the SFR is very low the choice of the adopted star--formation history (SFHs) in the fitting processes have an impact on the measured SFR itself. For instance, the popular exponentially declining star--formation histories do not reach a level of  $SFR\equiv0$ in finite time, and hence always yield a formally non--zero SFR. On the contrary, the  SED--fitting technique that we introduce in this work yields by definition a value of SFR$\equiv$$0$ (see below). To solve this ambiguity it is customary to set a threshold on the specific SFR (sSFR, the ratio between SFR and stellar mass) of  $<10^{-11}$ yr$^{-1}$ to identify galaxies where the SFR activity is effectively extremely low. We will use in the following the term ``passive'', or equivalently ``red and dead'', to identify such galaxies, that are the target of our analysis.
We will not use the term ``quiescent'' that is also often adopted, as it is indeed also utilized to identify galaxies that follow a non--starburst, gentle secular accretion of gas with a mild but non-null level of SFR.

Our data comes from a combination of the GOODS-S \citep{Giavalisco2004} and CANDELS \citep{Koekemoer2011, Grogin2011} surveys from the \textit{Hubble Space Telescope} (HST), the HUGS $K$-band survey \citep{Fontana2014} from VLT and a number of \textit{Spitzer} programs (see Sect. \ref{dataset}) that all conspire to make GOODS-South the best region of the sky where this kind of study can be executed over an area large enough to yield some statistically interesting result. We note in particular that the $K$ band is extremely useful since it straddles the 4000 {\AA} (rest-frame) break at $z=3-5$, a spectral feature of crucial importance to distinguish the (more abundant) star--forming dusty galaxies from the (rarer) passive sources \citep{Pozzetti2000}. Also, deep \textit{Spitzer} images are essential to disentangle the two populations, because of the different slope of the spectrum redward of $\sim$16000 {\AA} rest-frame, where dusty star forming objects have increasing fluxes while passively evolving galaxies have decreasing slopes. We use state-of-the-art software tools to extract reliable photometry from such low-resolution images (see Section \ref{dataset}). Finally, we complement this dataset with X--ray and far--infrared (FIR) catalogues by \citet{Xue2011}, \citet{Lutz2011}, \citet{Magnelli2013}, \citet{Smith2012} and \citet{Cappelluti2016}. Toward the end of our paper we will show how present-day best quality data are not sufficient to ensure unambiguous determinations of the properties of these high--redshift galaxies, and next generation telescopes will allow for a much more robust analysis.

However, there is another side of the story that has to be considered: one also has to check whether the data-analysis techniques adopted so far are well suited.
A typical approach is to select galaxies on the basis of their observed colours, e.g. in the $BzK$ plane at $z$$\simeq$2 \citep{Daddi2004}, or in its redward analogues at higher redshift \citep[e.g.][]{Guo2013}. Alternatively, the selection can be based on rest--frame colours, like in the widely adopted $UVJ$ diagram \citep{Labbe2005, Wuyts2007} in which the position on rest--frame  $V-J$ vs $U-V$ plane is considered. This technique has been used in up-to-date studies of galaxy formation \citep{Williams2009, Brammer2009, Patel2012}; recently, \citet[][ S14]{Straatman2014} used ZFOURGE data and the $UVJ$ colour selection to identify a substantial population of quiescent galaxies with $M_* > 10^{11} M_{\odot}$ at $z \sim 4$. A slightly different approach has been applied by \citet[][ N14]{Nayyeri2014}, who identified 16 post-starburst $z>3$ galaxies in GOODS-South, using the CANDELS photometry \citep{Guo2013} to perform a $Y-J$ vs. $H-K$ colour selection to probe the strength of the 4000 {\AA} break \footnote{We note that, contrary to the present study, both of these works do not strictly aim at singling out red and dead galaxies; they rather look for objects which have more generally quenched their major burst of star-formation, but allowing for a small residual ongoing activity.}.

A complementary approach consists in looking directly at the physical parameters resulting from the fit of the observed photometry by means of spectral energy distribution (SED) templates. Such models assume a SFH with an analytical form. A typical choice is an exponentially declining curve, with an $e$-folding time given by a parameter $\tau$, so that the SFR can be computed at any epoch $t$ as 
%$SFR(t) = (M_{final}/\tau) \times \exp[-(t-t_0)/\tau]$
$SFR(t) = SFR_0 \times \exp[-(t-t_0)/\tau]$. These are usually referred to as ``$\tau$-models'', and we will follow this convention here; the choice of such functional form was historically motivated by the need to model the quick formation and subsequent quiescent evolution of local ellipticals. 
Since in $\tau$-models the SFR formally never reaches zero, the threshold at which an object can be classified as passive is somewhat arbitrary. In previous analysis we adopted for instance a ratio age$/\tau>4$ \citep{Grazian2007}, or, as already said, a sSFR threshold $<10^{-11}$ yr$^{-1}$ \citep{Fontana2009}, which are in practice quite similar\footnote{The requirement sSFR$=10^{-11}$ yr$^{-1}$ corresponds to ages equal to 0.7 Gyrs if $\tau=0.1$ Gyr and 5 Gyrs if $\tau=1$ Gyr.}.
In other cases more physically motivated SFH have been adopted, for instance those extracted by theoretical hierarchical models \citep[e.g.][]{Pacifici2015}, and again a threshold sSFR$<10^{-11}$ yr$^{-1}$ has been used. 
%; however, to avoid confusion with the different modeling we have adopted in this study in which $\tau$ is a parameter (see below), 
% but here they will be referred to as ``ED models'' for the sake of clarity when making comparisons with other methods. 

It is now important to stress that the $UVJ$ and the SED--fitting methods are in practice basically equivalent, as the rest-frame $U, V$ and $J$ magnitudes that are used in the former, or at least the redshift estimation in absence of spectroscopic data, are in any case computed from SED--fitting, and hence subject to the same limitations and assumptions \citep[although it must be pointed out that there is much more information in a full multi-wavelength dataset than in only three bands, and in some cases the individual rest-frame $U$, $V$ and $J$ fluxes are derived by interpolating between observed bands; this is for example the case for \textsc{EAZY},][]{Brammer2008}. We shall discuss this equivalence in more details in the next Section. %, with an explicit mention to the popular $\tau$-models that are usually adopted for both.

What is instead crucial is the adoption of sensible SFHs for the underlying fitting. As it turns out, the adoption of $\tau$-models, while very successful at intermediate and low redshifts, cannot be the best choice at high-$z$: some galaxies would be excluded from a selection of passive sources because of the decaying exponential tail of SF in the best fitting models, which is an unavoidable numerical artifact. 
In fact, we will show in the following that this method is primarily sensitive to galaxies that are passively evolving since a relatively long time ($\geq 1$ Gyr), a requirement that is difficult (if not impossible) to match for galaxies at very high-$z$, given the small age of the Universe at such redshifts. 
Physically, this corresponds to excluding objects which have quenched their SF activity abruptly, and a short time before the epoch in which they are observed, for example because of gas stripping in dense environments, or strong energy feedback from galactic nuclei or young stellar populations. %This is of particular importance at high redshifts, where cosmic evolution takes place on short timescales.  
Therefore, in this paper we will adopt a different parametrization of the SFH, that is more suitable to select passive candidates at very high redshifts, as it assumes a single burst of SF activity abruptly quenched at some early times and followed by quiescence ever since. We will describe this method in detail in the next Sections. Of course, both descriptions are a clearly over-simplified parametrization of a previous history that is certainly more complex and irregular than our simple models. In principle, it would be good not to limit the fitting options to a single SFH shape, but explore different combinations of SFH.  In this study we choose to focus on the top-hat function as a good and simple prior to find galaxies that have been dead since some time at very high redshift, deferring to future work a more refined analysis.

A second aspect we will particularly focus upon is the reliability of the detection. At $z$>3 all the candidates selected so far are photometric candidates, singled out in deep extragalactic surveys with near--infrared (NIR) coverage. Because of their extreme faintness (they are typically fainter than $mag\simeq 26$ in the visible bands) a spectroscopic validation of their redshift and spectral classification is impossible with current instrumentation, and awaits \textit{James Webb Space Telescope} (\textit{JWST}) or \textit{Extremely Large Telescope} class instrumentation. Presently, at these faint magnitudes the combined effects of the limited number of available bands and the low signal--to--noise ratio (S/N) make the estimate of the photometric redshift and the spectral classification certainly difficult. %All the objects selected so far have been identified on the basis of the best-fitting SED template, that yields both the redshift and the spectral classification. 
In this work we have significantly extended the analysis of the reliability of the photometric selection, to investigate whether such candidates of passively evolving galaxies may be mis-classified star-forming dusty ones, taking into full account the photometric scatter and the corresponding redshift uncertainty.

The paper is structured as follows. In Section \ref{dataset} the analyzed dataset is presented. In Section \ref{UVJtheor} we discuss in more details the theoretical motivation for our particular approach based on truncated SFHs, and in Section \ref{method} the adopted selection criteria are described in details, together with the results of the selection processes. In Section \ref{others} we briefly discuss the differences between our results and the ones obtained using alternative approaches. %Section \ref{diagplanes} presents an analysis of our results based on some popular diagnostic planes. 
In Section \ref{jwst} we present some considerations on the future perspective with \textit{JWST}. Finally, in Section \ref{summary} summary and conclusions are presented. Tables with the physical properties of the selected objects, their multi-band snapshots and their SEDs are shown in the Appendixes.

All magnitudes are given in the AB system; we assume a standard concordance $\Lambda$-CDM cosmology with $H_0 = 70.0$, $\Omega_{\Lambda}=0.7$, $\Omega_m=0.3$.
%parameters derived from the Planck 2013 data release \citep[$H_0 = 67.15$, $\Omega_{\Lambda}=0.683$, $\Omega_m=0.317$;][]{Planck2014}. 

% %%% Fig 1
% \begin{figure}[h!] 
% \centering
% \includegraphics[width=8cm]{figs/tophat.png}
% \caption{ED model SFH (black line) and TH model SFH (red line). For both cases $\tau=100$ and $SFR_0=10$ (arbitrary units). If the "true" $SFR$ of the object was abruptly quenched around $t=100$, any decaying exponential model will inevitably include spurious SF activity in later times, leading to a non-passive fitting of the source.}\label{th}
% \end{figure}

\section{The dataset} \label{dataset}

%We use for this work the GOODS-South field data set. 
Most of the data used here have already been published in previous papers, and are based on the official CANDELS GOODS-South photometric and redshift catalogues \citep{Guo2013, Fontana2014, Dahlen2013, Santini2015}. The released photometric catalogue consists of 17 pass-bands, combining data from space and from ground, from ultra--violet (CTIO $U$ band) to mid--infrared (IRAC 8.0 $\mu$m). 

Since the release of this catalogue, new data have been acquired, and they are included in the present study. New VIMOS $B$ and HST WFC3 $F140W$ have been secured over the whole field, ensuring that the Lyman break is sampled for all the objects at $z\geq3$ (we note that the addition of the bands shortward of the Lyman-$\alpha$, that are in principle subject to further uncertainty because of stochastical intra--cluster medium absorption, do not affect our results, since our target galaxies are very red sources that are predicted to be well below the observed limits in the short wavelength bands, so that they are always consistent with the detected upper limits on the flux). Also, HUGS $Ks$--band data \citep{Fontana2014} have been made public. These new data had already been included in \citet{Grazian2015}.
We also use new, more accurate \textit{Spitzer} photometry, taking advantage of both new software tools and deeper images. For IRAC 3.6 and 4.5 $\mu$m pass-bands, we use a mosaic produced by R. McLure (priv. comm.), which combines images from seven observational programs \citep[Dickinson, van Dokkum, Labb\'e, Bouwens, and three by Fazio including SEDS and S-CANDELS: see][for details]{Ashby2015} into a single \textit{supermap}. The images are substantially equivalent to those recently released by \citet{Labbe2015}, and reach an average depth of $\sim25.7$ (total magnitude at $5 \sigma$) on both channels. On these new images we have used the code \textsc{t-phot} \citep{Merlin2015, Merlin2016} to derive the photometry starting from the same $H$-band detected objects of \citet[][G13]{Guo2013}, to which we added a sample of 173 $H$-undetected sources, detected in the $K$-band image (S/N>5; Boutsia et al., in preparation), and 5 IRAC-detected galaxies \citep{Wang2016}. Although improved in several aspects, \textsc{t-phot} is conceptually analogous to \textsc{TFIT} \citep{Laidler2007}, the code used to obtain the G13 catalogue; therefore, the new photometry can be cleanly combined with the previous catalog without introducing systematic effects. %For a full description of the methods and the images, see \citet{Merlin2016b}. 
IRAC 5.8 and 8.0 $\mu$m photometry is obtained again using \textsc{t-phot}, but on the CANDELS images. We have explicitly verified that the photometry of bright sources is statistically consistent in the new and old data set over all the IRAC bands, despite the adoption of the new images and tools.
This improved catalogue is an early version of a fully refurbished catalog of the GOODS-South fields that we plan to publish later (Fontana et al., in preparation). 
%The comparison between the new photometry and the CANDELS official one is shown in Fig.\ref{IRACcomp}.

\begin{figure*} %[t!] 
\centering
\includegraphics[width=18cm]{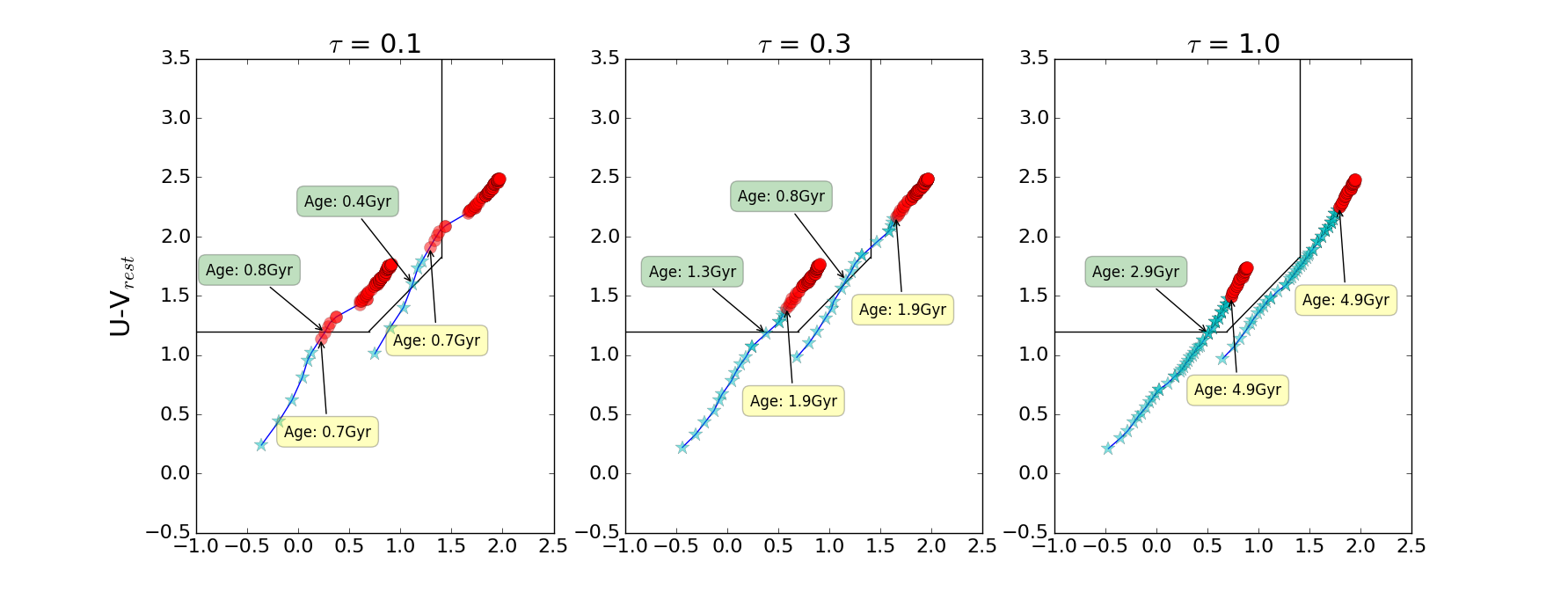}
\includegraphics[width=18cm]{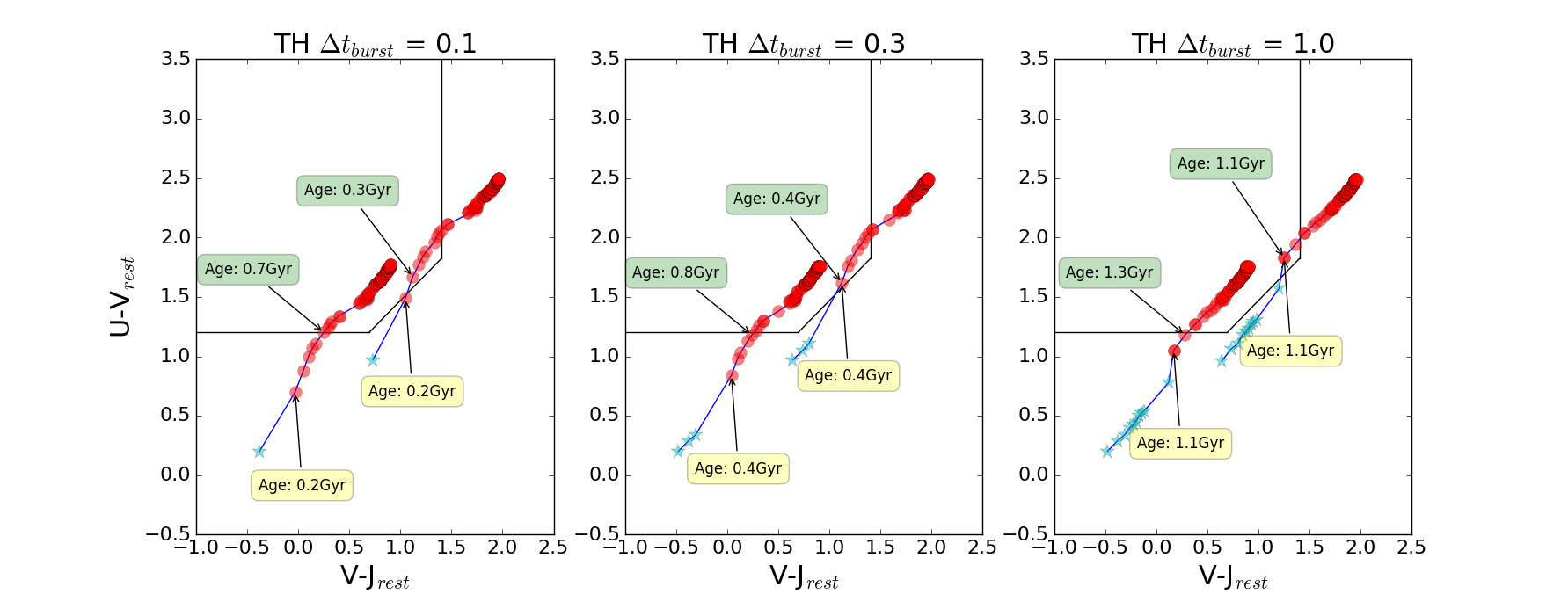}
\caption{Time evolution of model galaxies on the UVJ plane. Top: exponentially declining SFH ($\tau$-models); bottom: \textit{top-hat} SFH (TH models). Each panel shows a different timescale, as shown in the legends ($\tau$ is the $e$-folding time in the exponentially declining model, and $\Delta t_{burst}$ is  the duration of the burst in the top-hat models). In each panel, the star symbols refer to the star forming phase, and the red dots to the passive phase. Dots are placed in the diagram every 0.1 Gyr of evolution. Yellow labels indicate the age of the models when they become passive (i.e., sSFR$<10^{-11} yr^{-1}$). Green labels indicate the age when they enter the region of passive objects in the diagram (green labels).
In each panel, the leftmost trails refer to a galaxy with no dust extinction; in the rightmost models the extinction is equal to 0.4 (in principle, after the quenching of the SF activity the dust content is expected to decrease rapidly, so the rightmost trail should reconnect with the leftmost after some time). The aim of this figure is to show that a galaxy with a sudden truncation of SF (exemplified by the TH models) can  remain outside the UVJ selection region (and hence miss detection) for an extended period after the SF quenching. See text for more details.}\label{UVJteorico}
\end{figure*}

Even if we have obtained a revised photometry, we decided to adopt the photometric redshifts computed by the CANDELS team with the original G13 catalogue, and released in \citet{Santini2015}, as our baseline redshift estimate. We tested indeed the variation in photo-$z$ due to the new data, using our code $zphot$ \citep{Fontana2000} to obtain photometric redshift estimations on both the new and old catalogues. We found the differences to be small - unsurprisingly, given that the solution is dominated by the bands that have remained unchanged, which are more numerous and usually have higher S/N than those that have changed. Although our code performs well, the official redshift estimation provided by the CANDELS team is an optimal average of several photo-$z$ recipes, and we expect it to perform better than any individual technique; therefore, we decided that the advantages given by this overwhelm the possible disadvantages due to the adoption of a redshift estimate based on a slightly different photometry, and decided to keep the official CANDELS value. %We have verified that for our final selected sample of red and dead galaxies the differences in photo-$z$ are small.

Once the redshift has been fixed at the CANDELS one, we have however re-computed masses, SFRs and ages, as well as the other basic physical properties, again using $zphot$ and our improved photometry, as described in detail in Section~\ref{method}. 

At the end of the process, we have checked that our selection of passive objects has new best-fit redshift solutions broadly consistent with the CANDELS ones, and we finally performed a new SED--fitting run on the candidates, letting $z$ free to vary (see Section \ref{zfreesel}). 
%**** NON E' SEMPRE VERISSIMO, VEDERE FIGURA confronto\_z ****

%\section{A theoretical motivation: the \textit{UVJ} diagram selection} \label{UVJtheor}
\section{Tuning models for the search of high redshift passive galaxies} \label{UVJtheor}

As anticipated, a variety of techniques have been proposed and used in the literature to select passively evolving galaxies at high redshift. %They are either based on observed colours \citep[the most remarkable example is the $BzK$ diagram adopted at $z\simeq 2$, see][and modified redward to perform at higher redshifts - e.g. $VJL$ at $z\simeq 3$; see also Sect. \ref{diagplanes}]{Daddi2004}, on rest--frame colours (the $UVJ$ method mentioned above), or explicitly on the results of SED--fitting techniques. 
To some extent all these methods rely on spectral synthesis models, in two ways. First, these are used to define the selection area in the different colour planes adopted (either observed, as in the case of $BzK$ criterion, or rest--frame, as in the \textit{UVJ} one).
% In the following we will focus on the \textit{UVJ} diagram selection, which in recent years has become a widely used technique. Rest-frame colours are a powerful diagnostic to single out objects having no (or very little) ongoing star formation, at the same time excluding low-$z$ dusty starbursts: the net effect of dust extinction on the position of a galaxy on the \textit{UVJ} plane is a shift to the right side of the diagram, effectively leading it outside the region in which passive candidates are found. % As anticipated, we propose some modification to this technique, of which we describe the rationale here.
Second, SED modeling is at the basis of the techniques that use rest--frame properties of the studied objects, either colours or physical quantities like SFR, stellar mass and age. In particular, it is worth reminding that the cleanness of the \textit{UVJ} criterion stems from the adoption of SED modeling to estimate the rest-frame $U-V$ and $V-J$ colours of observed galaxies, which are either directly obtained from the best-fit models to the observed multi-wavelength fluxes distribution, or \citep[as in the case of the \textsc{EAZY} code,][]{Brammer2008} are computed interpolating between the observed bands but assuming a best-fit model to estimate the redshift of the object, unless spectroscopic data is available.

Given this importance of the SED--fitting procedure in the selection process, we dedicate this section to analyze whether the usual $\tau$--models are adequate, and to introduce a new simple parametrization that should be more effective at selecting passive galaxies at high-$z$.
% In most cases, the SED models adopted for both purposes are simple $\tau$-models, whose implications we describe in the following section.%, with SFR scaling as SFR$(t)\propto \exp{(-t/\tau)}$. %Strictly speaking, these model never reach SFR $\equiv 0$, although in practice they reach negligible level of residual star--formation rate if its timescale $\tau$ is much shorter than the age of the galaxy $t$ (typically when age$/\tau > 4$). 

\subsection{The $\tau$-models}

Because the SFR never reaches a value exactly zero in $\tau$--models (although in practice they reach negligible level of residual star--formation rate when the age of the galaxy $t$ is much larger than the timescale $\tau$), the exact definition of when a galaxy becomes passive in such models is somewhat arbitrary. In practice, a threshold on the specific star--formation rate is usually applied (e.g. sSFR$<10^{-11} yr^{-1}$). For this reason, in an attempt to define a safe area for passively evolving galaxies, both the (rest-frame) $UVJ$ and the (observed) $BzK$ diagrams adopt models in which the passive phase is well established and has been lasting for at least 1 Gyr, as we extensively discuss below \citep[see also][]{Sommariva2014}. These models can adequately describe the SFH of passively evolving galaxies even at redshifts $z\simeq 2-3$, with $\tau \leq 0.5$ Gyr; however, this requirement becomes difficult to match at high redshifts. Using the $UVJ$ diagram and adopting an exponentially declining model, one has to use very short $\tau$ parameters, often even below 0.1 Gyr combined with relatively large ages, to fit early passive objects (as clearly shown in S14, Table 1). Particularly for massive objects, which are the ones we can observe more easily, such extremely short timescales are difficult to reconcile with any physically-motivated description of the SFH at high redshift, and passive phases lasting more than 1 Gyr are difficult or impossible to hold at high redshift, when they become comparable to the age of the Universe (e.g., $\sim 1.2$ Gyr at $z\sim5$ in the $\Lambda$-CDM cosmology). 

This point is quantified in Fig. \ref{UVJteorico}. We use here the \textit{UVJ} plane to follow the colour evolution of model galaxies with different SF histories. The standard region used to select passively evolving galaxies is delimited by the solid lines, requiring $(U-V)_{rest} > 1.2$, $(V-J)_{rest} < 1.4$, and $(U-V)_{rest} > 0.88 \times (V-J)_{rest} + 0.59$ \citep[as defined by][for $z$$\sim$3 galaxies]{Whitaker2011}.

In the upper panels we show the path on the $UVJ$ diagram of galaxies following a $\tau$-model of SF with three timescales ($\tau$=0.1, 0.3 and 1.0 Gyr), short enough as necessary to describe quiescent galaxies at high redshift. 
Symbols are placed in the diagram every 0.1 Gyr of evolution; the labels indicate the age of the models when they become passive (i.e., sSFR$<10^{-11} yr^{-1}$; yellow labels) and when they enter the region of passive objects in the diagram (green labels). In all panels, the leftmost trail corresponds to a model with zero dust extinction, while the rightmost trail refers to a model with $E(B-V)$=0.4, described by the law by \citet[][ C00]{Calzetti2000}. Note that since the dust obscuration is only expected to play a major role mostly during the starburst phase, in reality the path of dust-obscured models would move towards the path of the unobscured models after the SF phase: therefore, even if the galaxy is strongly obscured during its first phases of life, it will most likely move to the region of dust-free models as soon as the SF process ends. It is clearly seen that even with very short $\tau$ the galaxies become passive %(i.e. with a ratio age$/\tau >4$ or with sSFR$<1\times 10^{-11}$yr$^{-1}$) 
at ages close to or higher than 1 Gyr, approximately when they also enter the $UVJ$ selection area. This long time required to reach the region is due to the tail induced by the exponential law. This age constraint may lead to miss some of the recently quenched objects at high redshift, where the time from Big Bang is comparable to the timescale needed to enter the $UVJ$ area \citep[the effects of not considering properly the abrupt quenching of the SF activity in starburst galaxies have been also recently discussed e.g. in][]{Ciesla2016}. It can be also noted that when $\tau$ increases to $\sim$1 Gyr, galaxies may enter the passive region of the diagram well before becoming quiescent, and stay there up to 2 Gyrs while still being star--forming (see the third panel).

%As anticipated, $\tau$-models are therefore good approximations for the SFH of quiescent galaxies at intermediate and low redshift, but may fail to reproduce objects where the SFR has been truncated by more drastic events and from relatively short time

To cope with this issues when searching for passive objects in the very early Universe, a different recipe might be more convenient.
%\textbf{We can conclude that the $UVJ$ diagram is not a reliable choice when searching for passive objects in the very early Universe. However, as we already pointed out, the $tau$-models themselves are not suited to models high--redshift objects. Therefore, before turning down the $UVJ$ selection technique it is worth exploring different, more convenient recipes.}

\subsection{The top-hat star--formation histories}

%%% Fig 11
\begin{figure} %[ht!]
\centering
\includegraphics[width=8cm]{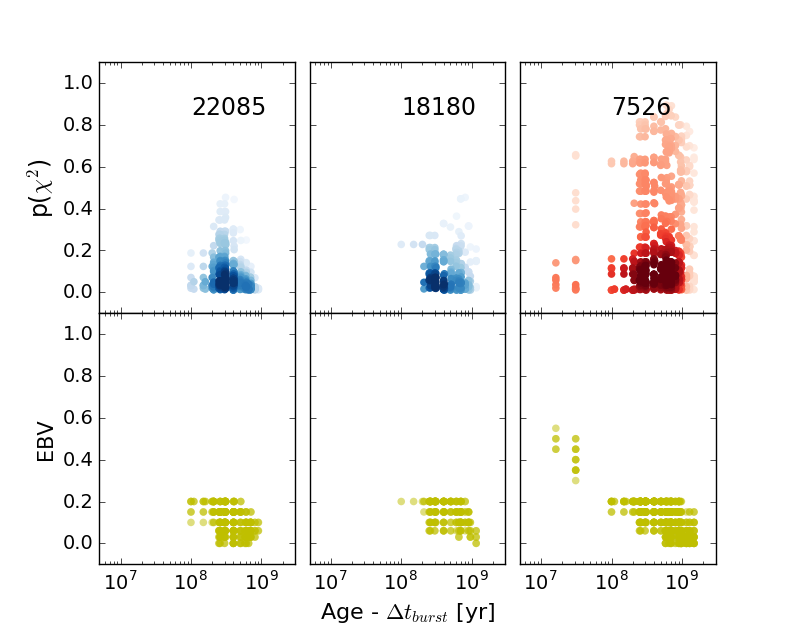}
\captionof{figure}{Probability ($p(\chi^2)$, upper panels) and $E(B-V)$ (lower panels) as a function of [age - $\Delta t_{burst}$] of all the possible solutions in the SED--fitting process, for three red and dead candidates in the reference sample (IDs 22085, 18180 and 7526, chosen as examples of three levels of robustness in the selection, see also Fig. \ref{zfreeseds}; the plot for the full TH sample is in Appendix \ref{probs}). The colours of the dots in the probability panels refer to the belonging of the source to the selection with (blue) or without (red) the inclusion of nebular lines. The dots (i.e. the models) are shaded as a function of their density. All the solutions have age $> \Delta t_{burst}$, as required to be classified as passive in this approach. Galaxies excluded from the selection, on the other hand, have been fitted by at least one model with age $\leq \Delta t_{burst}$, i.e. still star--forming, with a probability $p>5\%$ (not shown).}\label{probebvTH}
\end{figure}

Following the above line of reasoning, we complement $\tau$-models with another set of models that is (also) quite naive, but whose free parameters can be more indicative of the essential features of a truncated SFH. We define \textit{top-hat} (TH) SFHs those characterized by a constant star-formation rate for a time $\Delta t_{burst}$, after which the SFR instantaneously drops to zero. For ages $\leq \Delta t_{burst}$, the TH models are actively star forming, while for ages $> \Delta t_{burst}$ they are obviously evolving passively. 

These models have the obvious advantage of a clear transition between the star--forming, dusty phase and the passive one. Of course, the SFH of real galaxies in the star forming phase is definitely more complex, including random fluctuations on an overall rising or declining behavior; therefore, more realistic SFHs could have been used, e.g. including a rising SF phase before quenching occurs. Nevertheless, the very simple TH models do grasp the core of the matter as far as an abrupt and early quenching is considered, %. As already said, 
and even if more refined models would offer a better description of the galactic evolution in colours, this can be considered beyond the scope of this work. %In fact, after two galaxies quench their SF activity the differences in their colours due to different SFHs are expected to level out within a reasonably small amount of time (a few tens of Myr after the quenching), because young massive blue stars soon disappear in both galaxies, independently of the previous history; subsequently, most of the remaining stellar content is made of typical main sequence stars. To strengthen this assumption, 

In the bottom panels of Fig. \ref{UVJteorico} we show the outcome of our TH models ($\Delta t_{burst}$=0.1, 0.3 and 1.0 Gyr), that can be compared with the $\tau$--models of the upper panels. We recall that in this case the models are star--forming until the age is equal to $\Delta t_{burst}$, after which they are passively evolving. 
Two points can be highlighted. First, the colour tracks are always quite similar, also in comparison with the $\tau$-models (although the time needed to travel across the path is different): this implies that the details of the SFH before the quiescent phase do not affect significantly the position of the objects in the $UVJ$ diagram. %This can be understood considering that the colours of a passively evolving object basically depend on the aging of the main sequence stellar population, which was created during the burst, so they set to typical values soon after the death of the stars from the massive tail of the IMF (a few Myr), independently of the details of the previous SFH. 
More importantly, this parametrisation makes clear and unambiguous that galaxies enter the $UVJ$ selection area well after their actual quenching: %, with a time lag of up to 0.5 Gyr. In the colour space, the transition to quiescence can happen at a distance of $\sim$0.5 mag from the quiescent boundary.
% The models in the top panels all have $\tau=0.3$ Gyr; the ones in the bottom panels have $\tau=1$ Gyr ($\tau$ is the $e$-folding time in exponentially declining SFH models, and the duration of the burst in \textit{top-hat} SFH models; see also Fig. \ref{th}). 
%Looking at the path of the \textit{top-hat} objects i
%From this simple analysis, it is clear that 
a (dust-free) galaxy with abruptly quenched SFH can evolve passively for a long time (up to almost 0.5 Gyr) \emph{before} entering the selection region of the diagram. In the colour space, the transition to quiescence can happen at a distance of $\sim$0.5 mag from the quiescent boundary. As a consequence, the {\textit{UVJ}} diagram cannot be considered an optimal tool to select passive galaxies, at least at very high redshift. %, where such long times are not available.

%On the other hand, a fit to a recently quenched galaxy with an exponentially declining fit would make the object fitted as still active, with rest-frame colours that will classify it as star--forming. %This is particularly evident in the $\tau=0.3$ Gyr models. Because the Hubble time is still small at high redshifts, the burst of SF activity has to be short in terms of proper time, so the considered example is exactly the case that we face when searching with high-$z$ sources.

% We conclude that, albeit likely simplified, this modelization is useful to contain in a single set of models the two population that make red galaxies at high redshift.

Because of this, in the following we will primarily exploit the SED--fitting of the multi-wavelength photometry, rather than the simple position in the $UVJ$ diagram, and we will adopt the TH models as input. %In principle, we could re-cast the $UVJ$ diagram to be more sensitive to recently quenched objects (for instance by lowering the $U-V$ threshold), but we prefer to keep the standard formulation and to
Re-casting the $UVJ$ diagram to be more sensitive to recently quenched objects (for instance by lowering the $U-V$ threshold) would cause a specular problem making star--forming objects enter the passive region. Therefore, we prefer to directly use the information contained in the SED--fitting to characterize our objects.

\section{The selection of passive galaxies in GOODS-South} \label{method}

In this Section we describe the procedure we followed to identify the sample of red and dead candidates in our dataset, using the TH models described above.

\subsection{The SED--fitting method}

The SED--fitting has been performed on the 19 bands catalogue described in Section \ref{dataset} using our code $zphot$, in which we have implemented the TH models in addition to standard $\tau$-models. The same SED--fitting technique has been used in several previous studies \citep{Fontana2004, Fontana2006, Grazian2006, Maiolino2008, Santini2012, Dahlen2013, Castellano2014, Castellano2016} and it is similar to that adopted by other groups in the literature \citep[e.g.][]{Dickinson2003, Ilbert2013}; however, the adoption of the abruptly quenched SFH is novel.% The models adopted by our analysis are the \textit{top--hat} ones that we described before, that include very neatly both the star--forming and the quiescent phase. As we will show later they allow us to classify the two phases very simply using the output quantities of the fit.  

As described above, the TH library consists of a grid of models with constant star-formation rate for a time $\Delta t_{burst}$, after which the SFR is set to zero. The models have been created using \citet[][ BC03]{Bruzual2003} libraries and adopting a  \citet{Salpeter1959} IMF. Ages are computed from the onset of SFR, which means that any model is star--forming from age=0 to age=$\Delta t_{burst}$, and passive for age$>\Delta t_{burst}$. Only ages less than the age of the Universe at a given redshift are allowed. The burst duration $\Delta t_{burst}$ spans several values (0.1, 0.3, 0.6, 1.0, 2.0 and 3.0 Gyr) as well as  metallicities ($Z/Z_{\odot}$=0.2, 0.4, 1). For each value of $\Delta t_{burst}$, dust is included adopting C00 or Small Magellanic Cloud \citep{Prevot1984} attenuation curves, limited within the following physically motivated values:
\begin{itemize}
\item $0 <$ age $\leq \Delta t_{burst}$: $0 < E(B-V) \leq 1$
\item age $>\Delta t_{burst}$: 0 $< E(B-V) \leq 0.2$
%\item $\Delta t_{burst} <$ age $\leq \Delta t_{burst}$ + 0.3 Gyr: 0 $< E(B-V) \leq 0.2$
%\item age $>\Delta t_{burst} + 0.3$ Gyr: $E(B-V)=0$
\end{itemize}

\noindent (this choice mimics the expected drop of dust content in a quenched galaxy after the end of the star forming activity). The full library consists of $\sim$3.13 millions models, and the quasi-logarithmic step in age results in a larger number of star--forming models ($\sim$78\% indeed have age$<\Delta t_{burst}$).

%Summarizing, we will fit the galaxies in the CANDELS data set using our TH models. 
Another important ingredient in computing the SED of high-$z$ galaxies is the proper inclusion of emission lines, that can contribute significantly to the observed $K$-band and IRAC fluxes \citep[e.g.][]{Nayyeri2014, Pacifici2015}. 
As first presented in \citet{Castellano2014},  the contribution from nebular emission has been inserted in $zphot$  following \citet{Schaerer2009}. Briefly, nebular emission is directly linked to the amount of hydrogen-ionizing photons in the stellar SED \citep{Schaerer1998} assuming an escape fraction $f_{esc}=0.0$. The ionizing radiation is converted in nebular continuum emission considering free-free, free-bound, and H two-photon continuum emission, assuming an electron temperature $T_e=10 000$ K, an electron density $N_e=100$ cm$^{-3}$, and a 10\% helium numerical abundance relative to  hydrogen. Hydrogen lines from the Lyman to the Brackett series are included considering case B recombination, while the relative line intensities of He and metals as a function of metallicity are taken from \citet{Anders2003}.

However, the computation from first principles of this contribution is not easy, and it has not been tested on large spectroscopic samples. For this reason we choose to adopt two different paths: we build models both without emission lines, as done in most of the published analysis so far, as well as including emission lines, as described above. We will analyze our sample separately with both libraries.

The crucial output parameters of the fit are the SFR, which must be equal to zero in passive candidates, and (equivalently) the galaxy age, that must be compared with the duration of the SF activity: for ages shorter than $\Delta t_{burst}$, the galaxy is star--forming - these models can be described reasonably well both as starburst galaxies with a relatively small amount of dust (i.e. the usual Lyman Break Galaxies) as well as more reddened galaxies - while for ages definitely larger than $\Delta t_{burst}$ these models describe galaxies that are passively evolving, with negligible amount of star--formation activity.

\subsection{The selection criteria}

%As described in Sect. \ref{dataset}, 
We performed our search for passive galaxies in the GOODS-South field starting from the  $H$-detected catalogue, and using the photometric data described in Sect. \ref{dataset}.
First of all, we selected all the sources in G13 having $H160<$27, with simultaneous 1 $\sigma$ detection in \textit{Ks} (Hawk-I), IRAC 3.6 and IRAC 4.5 $\mu$m bands. We also excluded from the selection any source with defects or unclear classification (relying on the CANDELS flagging). Finally, we added the $K$/IRAC-detected sources to the list. %, or likely to be AGN from a spatial cross-correlation with the X-ray detected catalogues by \citet{Cappelluti2016} and \citet{Xue2011}. %QUI DOVREMMO USARE CAPPELLUTI.. ALMENO NOI!
On this sample, we have performed the SED--fitting process described above.

\begin{figure}
\centering
\includegraphics[width=8cm]{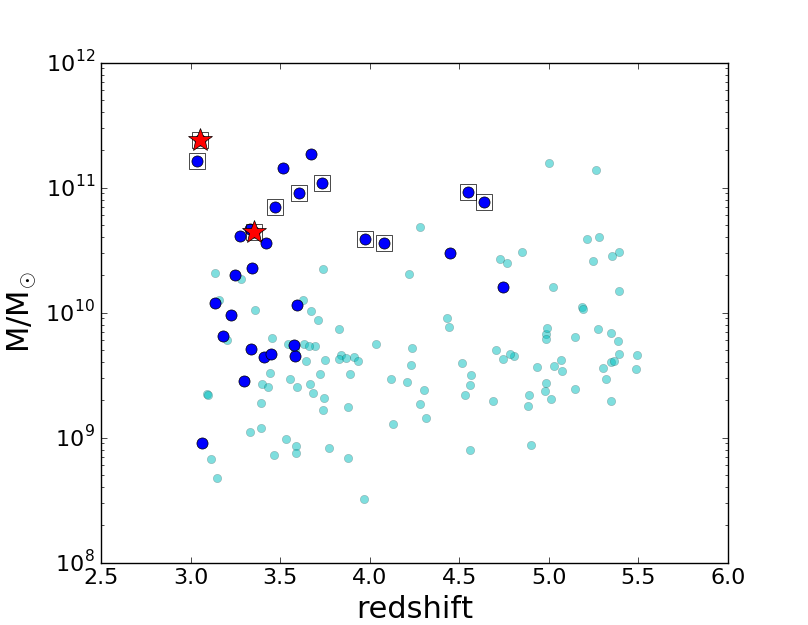}
\caption{Redshift vs. mass distribution of the red and dead selections. Cyan points refer to  objects having $z$$<$5.5 and a passive best--fitting solution (both in the fit including  emission lines and in the one without them), but excluded from the reference sample because of the existence of different, non--passive solutions with probability larger than 5\%; blue dots: reference sample (30 candidates surviving the probabilistic selection; the void squares mark the 10 sources selected also using the library including nebular lines emission); red stars: the two candidates surviving the free-$z$ fitting. See text for more details.} \label{z_distr}
\end{figure}

The selection is done using the information contained in the probability $p(\chi^2)$ of any fitted model. The probability $p(\chi^2)$ is simply computed as the probability that the observed $\chi^2$ (computed on all bands) is due to normally distributed errors on all the observed bands (the number of free parameters used in the calculation is actually $N-1$, with $N$ the available number of bands for each object, as one degree of freedom is used to normalize the spectrum). Similarly to what has been done in several previous works \citep[e.g.][]{Papovich2001,Fontana2009,Santini2015}, the procedure to estimate the $\chi^2$ is iterative. The $\chi^2$ of each object is first evaluated on its photometry, and then the uncertainties on the observed bands are all increased by a constant factor, in order to have the $\chi^2$ of the best-fitting solution equal to 1. This boost of the photometric uncertainties (that does not change the choice of the best-fitting solution) is used to take into account the limitations of the template models, that may lead to relatively large $\chi^2$ (the average $\chi^2$ is indeed $\simeq 3$) despite an overall good fit. We note that this approach is particularly conservative, as it widens significantly the allowed parameter space, and hence the possibility that a given object is classified both as passive and star--forming. Uncertainties on all the other relevant physical quantities in the fit are computed in the same way.

Our proposed method use the probability $p(\chi^2)$ by requiring that the following two conditions are fulfilled:
\begin{itemize}
\item no star--forming solutions with a probability higher than a fixed threshold $p_{SF}$ exist;
\item the best-fitting solution is characterized by an age $t$ larger than the duration of the burst $\Delta t_{burst}$, and must have a probability $p\geq30\%$.
\end{itemize}

%While the former is an obvious necessary condition, the latter 
The first of these two criteria enforces the reliability and credibility of the candidates by requiring that no plausible star--forming solution alternative to the best--fit one exists; in practice, this is obtained by looking at the probability $p(\chi^2)$ of every model inside our multi--parameter grid (that includes dusty--star forming models as well as passive ones). We explicitly note that our criterion is based on the absolute value of the probabilty $p(\chi^2)$, not on the density of the resulting models in the parameter space, as the latter would be indeed heavily altered by the (arbitrary) sampling of the models along the (many) free parameters.

To assess the value of $p_{SF}$, we used a set of dedicated simulations, proceeding as follows. We created a mock catalogue using the TH library, consisting of $\sim$2500 star--forming models having $H$ magnitudes of 23, 24 and 25. Each of these models was then replicated 10 times adding observational noise, consistently with the scatter of the distribution observed in each CANDELS pass-band; the full catalogue therefore consisted of $\sim$25000 mock objects . Then, we used the TH library to fit these models. Around 1200 of them turned out to have a passive best--fit, showing that the effect of noise and models degeneracy can turn a star--forming observed source into a passive fit, with some 5\% of chances. Our goal was therefore to make sure that the chosen criteria are stringent enough to avoid that any of these sources is confirmed as passive after the probabilistic selection. Unsurprisingly, these objects also had a non-zero probability of being fitted with a star--forming solution, although with a worst $\chi^2$ than the passive best-fit one; we found that the lowest probability of a star--forming solution is 12\%. Considering that the simulation is an idealized case and in real photometric catalogues the uncertainties due to blending or varying depth must be taken into account, we decided to apply a more conservative threshold of  $p_{SF}=$5\%.

As we show in the following, the effects of this first condition depend dramatically on the range of redshifts allowed in the error analysis. On the other hand, the second condition is significantly affected by the inclusion of the emission line in the spectral library. 
We list below the different samples resulting from the application of these alternatives.

\subsubsection{The selected reference sample}

The first fit has been performed keeping the redshift fixed at the CANDELS value \citep[i.e., spectroscopic redshifts when available, photometric redshifts otherwise, the latter obtained as a median of 9 independent photo-$z$'s determinations; see][]{Dahlen2013}. For all our objects, this corresponds to the photometric redshift, as they are too faint to be observed spectroscopically.

We therefore apply our SED--fitting technique to fit all the objects at $z_{CANDELS}$>3, utilizing the TH models \textit{without emission lines} and applying the selection criteria described above. 

This way, we single out 30 red and dead candidates. This approach is analogous to previous works \citep[e.g.][]{Fontana2009}: we use models with no emission lines and perform the scan of star--forming alternative solutions only at the best--fitting photometric redshifts. For these reasons, in the following we will consider this set of objects our \textit{reference sample}. These objects span a redshift range between 3.0 and 4.7, and have IRAC magnitudes 23$<m_{4.5}<$26 (corresponding to stellar masses between $10^9$ and $2 \times 10^{11}$ M$_{\odot}$ in our fit). Five of them have $z_{CANDELS}>4$ (IDs 3912, 5592, 6407, 9209, 23626); these all have best fit masses larger than $10^{10} M_{\odot}$. 
The 30 selected objects are $H$-detected (while none of the $K$/IRAC-detected additional sources passed our selection criteria). A full description of their physical best--fit parameters is shown in Appendix A, along with their images in the CANDELS data set (Appendix B) and full SED and resulting best--fitting spectrum (Appendix C). 

It is important to remark that the condition that the probability of star-forming solutions is less than $p_{SF}=5\%$ has a dramatic effect of the selection of candidates. Indeed, the number of objects that have a formal best-fit passive solution is much larger - namely 482 candidates ($1.4\%$ of the whole G13 catalogue, and $9.4\%$ of all $z$>3 galaxies). Most of these candidates are of course faint sources with detections at very few $\sigma$ levels in the $K$ and IRAC bands whose fit is degenerate, i.e. that can be fitted nearly equally well by star--forming or passive solutions. Indeed, almost all of the objects in the final reference sample have observed $K$, IRAC 3.6 and 4.5 magnitudes <25, with S/N respectively larger than 20, 10 and 6. Clearly, a robust analysis is possible only for galaxies well above the detection limit. This first example highlights the importance of the S/N in the credibility of the identification, on which we shall expand below.

In Fig. \ref{probebvTH} we show the outcome of this procedure for three objects belonging to the reference sample (with additional information, to be explained in the next Section); the whole sample is shown in Appendix \ref{probs}. For each object, we plot the probabilities of all the SED--fitting solutions (shaded as a function of their density), and the corresponding UV extinctions, as a function of the time passed from the end of the SF burst; the candidates have solutions with high probability and low extinction well after the quenching of the activity. The extinction tends to anti-correlate with time, because of the way our models are built, but also because a red object can be fitted with a young dusty model or with an old model without dust, so the two possibilities are somewhat degenerate and only the goodness of the fit (namely the $\chi^2$) can disentangle them.

\subsubsection{Including the emission lines}

%%% Fig 11
\begin{figure}
\centering
\includegraphics[width=8cm]{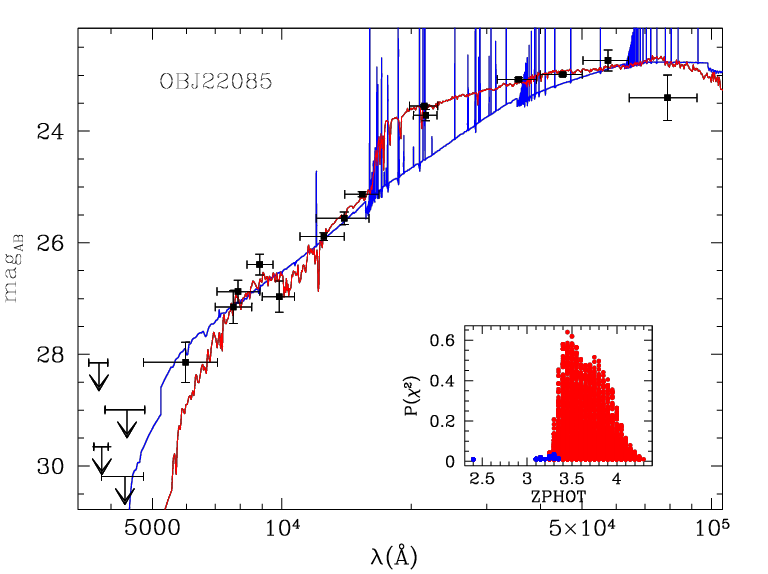}
\includegraphics[width=8cm]{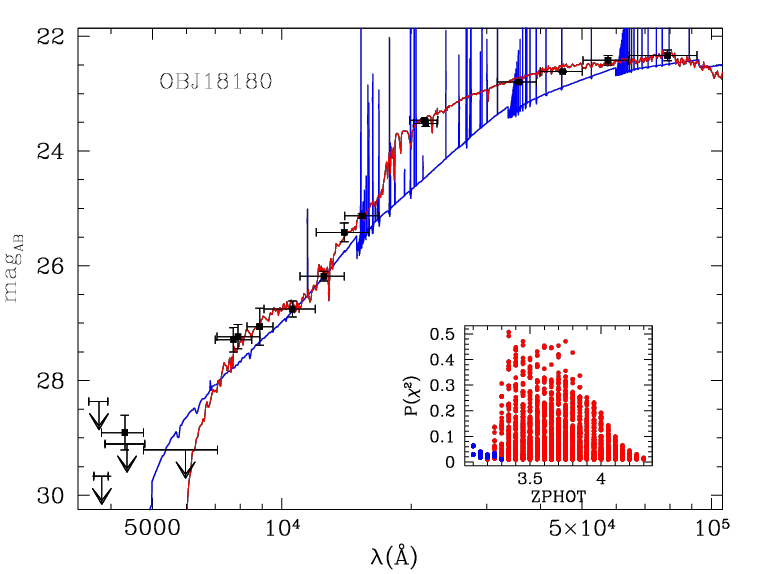}
\includegraphics[width=8cm]{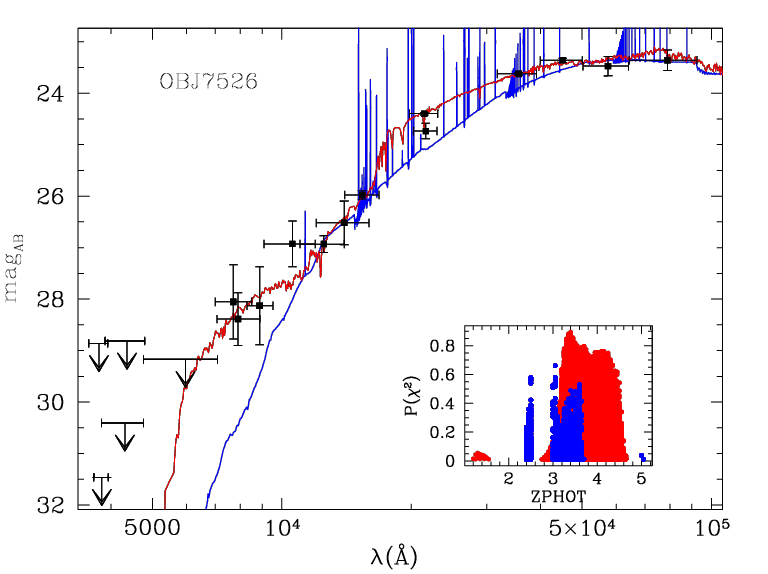}
\captionof{figure}{Fitted SEDs of the three candidates reported in Fig.~\ref{probebvTH} in the reference sample, and compared probabilities of all the solutions (inner panel), in the free-$z$ fit. In each panel, the red line is the best passive solution at the CANDELS photometric redshift, and the blue line is the best star--forming solution (at a different redshift); the dots in the inner panel refer to the corresponding probabilities. Top: ID22085, a strong candidate surpassing all the selection criteria: the best star--forming solution (obtained at $z$$\simeq$3.4) has probability $p$<5\%, so this object can safely be considered passive even letting its redshift freely vary in the fit. Middle: ID18180, despite having mostly passive solutions with very high probabilities, it also has a few star--forming solutions with $p$>5\% (at $z $$\simeq$3.1), so it is formally excluded from the most stringent selection; Bottom: ID7526 (which is part of the reference sample, but fails the emission line fit selection) has many high--probability star--forming solutions at various redshifts.}\label{zfreeseds}
%SEDs of one the best passive candidates, IDs 22085, and compared probabilities of all the solutions (inner panel), in the free-$z$ fit. The red line is the best passive solution, obtained at $z$$\simeq$3.45 ($z_{CANDELS}$ is 3.36); the red dots in the inner panel refer to the probabilities of all the passive solutions. The blue fit is the best star--forming solution, at $z$$\simeq$3.4: this solution has probability $p$<5\%, as shown by the blue dots in the inner panel.}\label{zfree22085}
\end{figure}

%{\bf TBD: Expand figure 4 in order to better show the various cases of fits}

We then repeated the analysis, but this time using the TH spectral models \textit{including} emission lines. 

In this case, we identify only 10 objects satisfying the selection criteria. As in the previous case, the requirement on the (low) probability of the star--forming solution is very effective in removing potential candidates at low S/N (we identify a total of 194 candidates with passive best-fit solutions). 

We remind here that the lines are computed self--consistently from the intensity of the ionizing flux in the spectrum, such that the resulting SED of a quiescent galaxy is by default identical to that obtained without emission lines. As a consequence, all the objects selected as passive in this ``emission lines sample'' are also part of the reference sample by construction.

On the other hand, of the 20 candidates of the reference sample that are not included in the emission lines sample, 12 have a star--forming best--fit solution with emission lines (i.e. best--fit ages $<\Delta_{burst}$) with significant star--formation rates (typically 30-50 M$_\odot$/yr). The physical origin of this difference can be understood looking at the SEDs showed in the Appendix C. The intense emission lines significantly affect the shape of the spectrum (as sampled by the broad band filters) beyond 4000 {\AA}, implying a weaker break and yielding a shallower slope in the rest--frame infrared region.

The remaining 8 objects still have a best--fit passive solution, but also an increased probability of having a star--forming solution, and hence do not pass our probabilistic selection criterion $p<5\%$ (see Table in Appendix A).

Looking at Fig. \ref{probebvTHall}, one can see how in the emission lines sample (blue dots) four sources are likely to be passive since more than 100 Myr (IDs 2782, 18180, 22085, 23626), while the other six might have quenched their SF activity very recently. In the reference sample (red dots), most of the sources have solutions suggesting very recent quenching, while only four are more likely to be passive since more than 100 Myr (IDs 3973, 4503, 7526, 7688). This is interesting particularly when compared to the results of a standard selection using decaying SFH histories (see Sect. \ref{others}). 

Fig. \ref{z_distr} shows the stellar masses and redshifts of the selected candidates. The emission lines sample sources (empty squares) are among the most massive objects in the reference selection (blue dots). This may be expected because the ambiguity between the passive and star--forming solution is increased by the somewhat lower S/N of the faintest among the 30 objects in the reference sample.

In principle, spectral templates including emission lines are expected to be a more accurate representation of the real spectra, and the intensity of such lines are expected to increase at high redshift, so that the 10 objects that are classified as passive even including the emission lines should be regared as more reliable. However, the recipes adopted here to compute the emission lines from the SED have not been extensively tested or verified, especially in high redshift galaxies. In addition, most previous works adopted spectral models without emission lines, so that a proper comparison can be more appropriately performed considering the whole reference sample. Therefore, we keep it as our fiducial selection.

\subsubsection {Free redshift selection} \label{zfreesel}

As already pointed out, the analysis described above assumes that the redshift is fixed at the CANDELS photometric estimate. However, for galaxies that have a very steep spectrum and are undetected in many of the optical images, the possibility of degeneracies among the spectral templates due to low S/N and poor sampling, and/or by the adoption of incorrect templates, may lead to substantial uncertainties in the photo-$z$ that need to be taken into account.
To this aim, we repeat again the SED--fitting procedure on the reference sample, but leaving the redshift free to vary around the best-fitting CANDELS value in the whole redshift range where the probability $p(z)$ of having an acceptable fit is above 1\%. 

We note that, since the best-fit photometric redshift has not been computed with the TH library, it is in principle possible that the best fitting photometric redshift obtained using the TH library is different from the official CANDELS one. 
Reassuringly, we find that most of the candidates still have best fit solutions with zero star formation activity, at redshifts similar to the CANDELS one. 

However, most galaxies also have star--forming solutions at different (typically lower) redshifts, with $p>5\%$. Only two galaxies, IDs 10578 and 22085, are left as reliable passive candidates, with no probable star forming solutions at any redshift. 

Of course, this does not mean that no other candidate is reliable as a ``real'' passive object. Indeed, the consistency between the best fit solutions leaving $z$ free to vary, and the one obtained at $z_{CANDELS}$ is reassuring; furthermore, we will show in the next Section that most of the objects in the reference sample have no detectable FIR emission, ensuring that most of them can be considered as robust red and dead candidates. Nevertheless, using the present-day state-of-the-art facilities and methods it is still not possible to rule out the possibility that some low-redshift, dust-reddened star-forming objects are erroneously identified as high-redshift passive galaxies with 100\% certainty. The best we can do is try and reduce the risk of contamination using all the available information, e.g. checking the FIR fluxes, while waiting for even deeper data to come (see Section \ref{jwst}).

As a summary of the $z$-free selection and a final summary of the whole procedure, Fig. \ref{zfreeseds} shows the fitted SEDs of the same three sources already discussed in Fig. \ref{probebvTH}, which passed three different levels of our selection criteria: top to bottom, ID22085, which survived the $z$-free selection and is one of the two strongest red and dead candidates; ID18180, which passed the emission line selection but has a star--forming solution at a redshift different from the CANDELS one; and ID7526, a ``standard'' object in the reference sample. In each panel, we plot the best fitting model at $z=z_{CANDELS}$ (which is always passive) as a red line, along with the best star--forming model at any redshift (blue line). The corresponding probability distributions can be inferred from the inner boxes.

\subsection{FIR fluxes}

To further reduce the risk of including dust-obscured star--forming solutions in the reference selection, we perform a sanity check on the \textit{Herschel} images and catalogues described in \citet{Magnelli2013}. The FIR images are shallower than the optical and NIR ones, but they can nevertheless provide a hint on the real nature of these objects; and indeed, detections in \textit{Herschel} bands occurs in many cases of passive candidates detected with other methods (see Section \ref{others}).

We first perform a spatial cross-correlation between the $H$-band coordinates of the passive candidates and the 24 $\mu$m MIPS catalogue. We find that 2 among them, IDs 3973 and 10578, have very close counterparts (below 1"), while none of the remaining 28 have one within a radius of 3.0" (the FWHM of MIPS is $\sim$5.7", but the catalogue has been obtained using IRAC 3.6 $\mu$m priors with FWHM $\sim$1.6", so this minimum distance is enough to exclude the detection of 24 $\mu$m flux). We note that ID10578 is indeed one of the ``strongest'' candidates in our reference sample, since it has survived the whole selection process, including the free-$z$ SED--fitting - meaning that no star--forming solution exists at any redshift with probability above 5\%.
Emission is detected at the position of the two sources also at longer wavelengths (100, 160 and 250 $\mu$m) on the \textit{Herschel} PEP-GOODS \citep{Lutz2011} and HerMES \citep{Smith2012} blind catalogues; however, possible associations become more common, given the increasing width of the PSFs: at 100 $\mu$m four sources have matches below the FWHM of $\sim$6.7", at 160 $\mu$m eleven sources have matches below the FWHM of $\sim$11.0", and at 250 $\mu$m sixteen objects have a match below the FWHM of $\sim$18.1". A visual inspection on the Herschel maps always hints at different possible objects as the origin of the detected fluxes, as many other $H$-detected galaxies lie close to the considered \textit{Herschel} source, so that it is almost impossible to discern the actual origin of the FIR emission. % (see Fig. \ref{herschel} for two examples).

To further strengthen the analysis, we have also checked a new \textsc{Astrodeep} MIPS/\textit{Herschel} catalogue by \citet{Wang2016}, which is deeper than previous catalogues particularly in the SPIRE bands, and uses the G13 $H$-band detections as priors. Using this catalogue it is therefore possible to directly link each source to its measured FIR flux. The two candidates identified above as having clear MIPS/\textit{Herschel} counterparts are also recognized as FIR emitters in this new catalogue, while none of the other red and dead candidates is associated with a detectable \textit{Herschel} source.

Finally, we checked a stack of the 28 non-associated sources thumbnails from the \textit{Herschel} maps, finding no trace of detectable flux in any of the considered bands.

Summarizing, it is fair to conclude that there is no emission at $\lambda \geq 24 \mu$m evidently linked to 28 out of 30 sources in the reference sample. Two objects instead clearly have FIR counterparts. In principle, this might be due to their wrong identification as passive in the SED--fitting process. However, as discussed above, the uncertainty in the classification dramatically worsens for objects with low S/N, while both the two galaxies are very bright (ID 3973 has S/N$_{H}$=17.3 and S/N$_{K}$=49.9, and ID 10578 has S/N$_{H}$=166.8 and S/N$_{K}$=528.9). A possible different explanation for their strong FIR emission might be the presence of a dust-obscured AGN hosted in a recently passivized galaxy. In this case, while the stellar content would yield a passive spectrum in the optical and NIR wavelengths, galactic dust absorbing and re-emitting part of the X radiation from the nuclei could cause the observed \textit{Herschel} fluxes. To test this hypothesis, we check the \textit{Chandra} catalogue by \citet{Cappelluti2016}, which directly links X--ray emitters to the $H$--detected sources in G13. As it turns out, both the two sources are identified as X--ray emitters in the catalogue. Also, in our analysis they are fitted with a relatively high amount of extinction ($E(B-V)$=0.2 for ID3973 and 0.3 for ID10578). This leads to the interesting speculation that these two galaxies indeed seem to have recently become passive, but still retain an active radiating nuclues, and large amounts of dust which cause the observed FIR strong emission. For these reasons, we decide to keep the two sources in the reference sample, although we are aware that red colours in the NIR/MIR wavelength range can be also typical of AGNs hosted in young, active galaxies \citep[see e.g.][]{Giallongo2015}, while our SED libraries are solely based on stellar tracks.
%A more detailed analysis would be necessary to completely rule out the possibility that any of our candidates is responsible for significant FIR emission, but this goes beyond the scope of this study. We feel fair to conclude that there is no obvious detected emission at $\lambda \geq 24 \mu$m evidently linked to any of the reference sample candidates.

\subsection{Properties of the red and dead candidates}\label{properties}

\begin{figure}
\centering
\includegraphics[width=8cm]{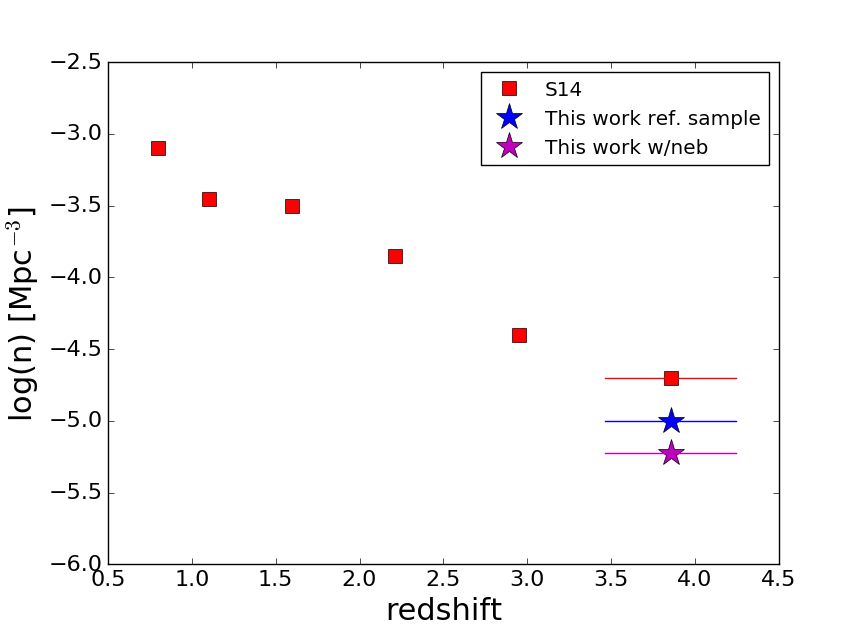}
\caption{Number density of red and dead candidates with log[M/M$_{\odot}$]>10.6 in the redshift bin 0.65<log(1+$z$)<0.72, from this work (blue star: reference sample; magenta star: ``emission lines'' selection), compared to the results by S14 (red squares). Our mass estimates are corrected to match the analysis by S14 who assumed a Chabrier rather than a Salpeter IMF. %The fainter stars refer to the total number density computed including candidates of all masses, and are shown to demonstrate how the fit including emission lines mostly select massive objects. 
See text for more details.} \label{ndens}
\end{figure}

%The relevant fitted physical properties of the galaxies in the reference sample are listed in Appendix A. The table starts from the most robust candidate (i.e. the only galaxy that ``survived'' the whole selection process), continuing with the other eight belonging to the selection obtained including the emission lines, and finally passing to the rest of the reference sample.

Four galaxies in the reference sample have been observed spectroscopically: IDs 4503 (Mobasher, priv. comm.), 9209 \citep{Cassata2015}, 10578 \citep{Vanzella2008}, 19505 (VANDELS, but the observations and analysis are not completed yet). In all the cases, $z_{spec}$ is close to $z_{CANDELS}$, except for 10578 ($z_{spec}=3.89, z_{CANDELS}=3.06$). However, all of these spectra have poor quality flags, so that they cannot be taken as strong constraints to our aims; the only exception might be ID19505, which shows a broad line at $\simeq 5640$ {\AA} that can be interpreted as a strong  Lyman-$\alpha$ emission. This would imply $z_{spec}$=3.6386; fitting the observed photometric data at this redshift yields a SED which is very similar to the one at $z_{CANDELS}$=3.33, but with a worse $\chi^2$, which would exclude the object from the selection.

Considering the area of the GOODS-South deep field ($\simeq$173 arcmin$^2$), the 30 passive candidates would imply a number density of $\sim$0.173 passive objects per arcmin$^{-2}$, at $z$>3 and above the detection criteria. The corresponding total comoving number density in the redshift interval 3<$z$<5 is of $\sim$$2.0\times10^{-5}$ Mpc$^{-3}$. Fig. \ref{ndens} shows a comparison between the number density of passive high redshift objects inferred from these study and the one found by S14: applying their same mass selection criterion of M>$10^{10.6}$ M$_{\odot}$\footnote{We apply a scaling factor of 0.24 dex \citep[e.g][]{Santini2012} to take into account the fact that we adopt a Salpeter IMF rather than a Chabrier IMF as in S14.}, and considering their redshift bin 0.65<log(1+$z$)<0.72, we find a number density of $\sim$$1.0\times10^{-5}$ if we consider the whole reference sample, and of $\sim$6.0$\times10^{-6}$ if we only include the emission lines selection. These densities are slightly lower than the $\sim$$1.78\times10^{-5}$ value found by S14 - which is unsurprising given their more relaxed selection criteria - and are broadly consistent with the value found by \citet{Muzzin2013}.

\section{Comparison with other selection criteria} \label{others}

%%% Fig 
\begin{figure}
\centering
\includegraphics[width=9cm]{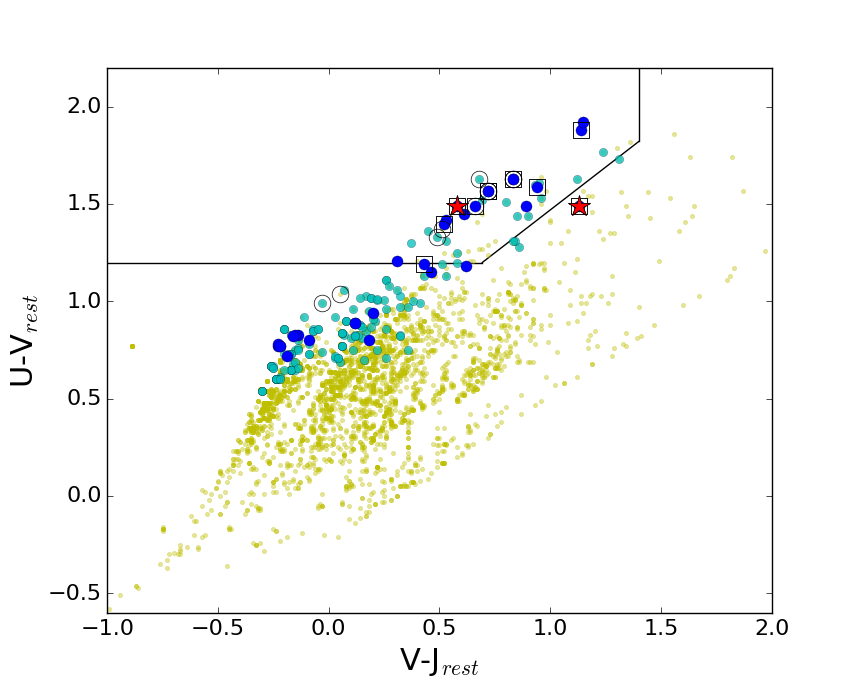}
\caption{$UVJ$ diagram for $z_{CANDELS}$>3 sources, computed adopting the standard $\tau$-models to derive the rest-frame magnitudes. The passive selection region is defined by $(U-V) > 0.88 \times (V-J)+0.59$, $(U-V)>1.2$ and $(V-J)<1.4$ \citep{Whitaker2011}.  
Yellow points are the whole sample of $z$>3, $Ks$+3.6+4.5 $\mu$m $H$-detected galaxies from G13. Cyan small dots refer to the objects having a passive best--fitting model both with and without the inclusion of the emission lines in the TH library. Blue large dots are the 30 galaxies in the reference sample, with empty squares indicating the 10 sources selected also including nebular emission lines. The red stars refer to the candidates surviving the $z$-free selection. Finally, empty circles refer to the $\tau$-model red and dead selection (sSFR $< 10^{-11}$ yr$^{-1}$, Sect. \ref{others}). Not all the galaxies in the selections are visible because some models have very similar rest-frame properties and colours, so that the symbols overlap.} \label{uvj}
\end{figure}

%%% Fig 
\begin{figure}
\centering
\includegraphics[width=9cm]{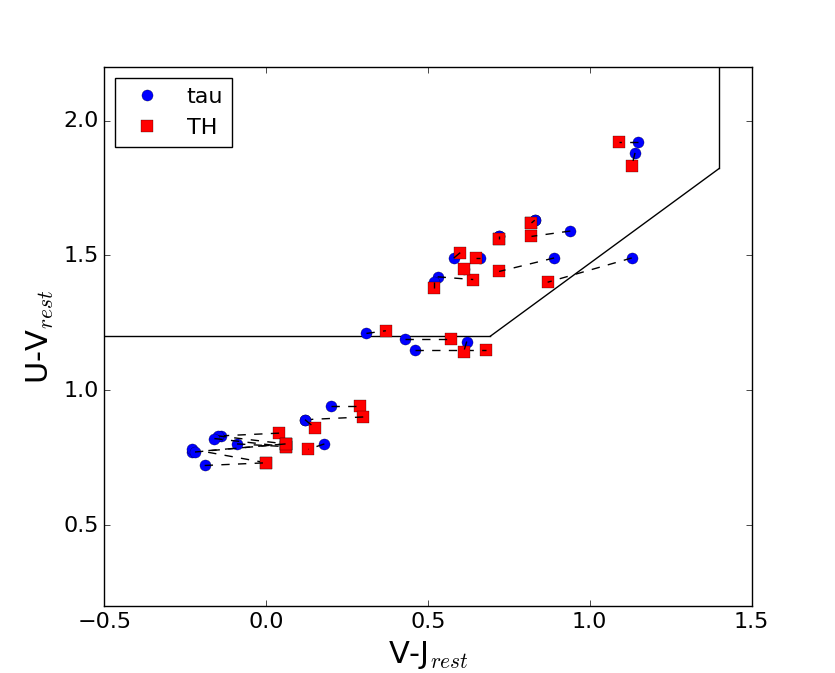}
\includegraphics[width=4.1cm]{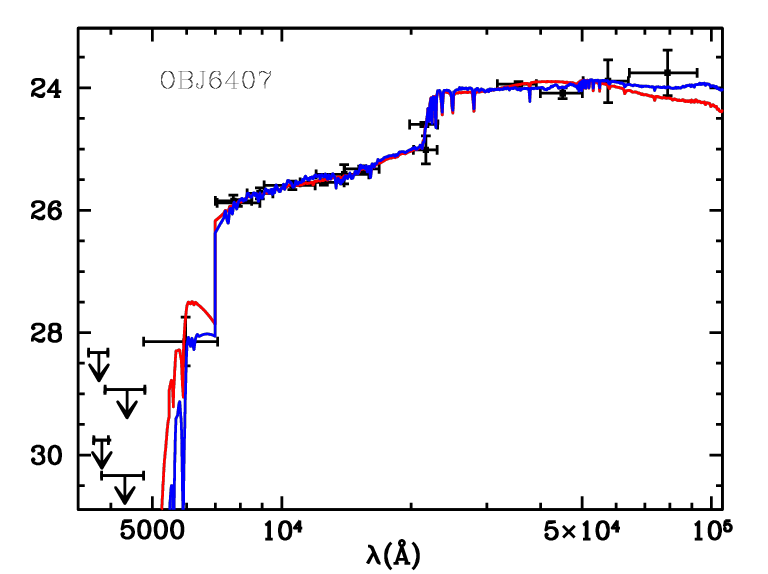}
\includegraphics[width=4.1cm]{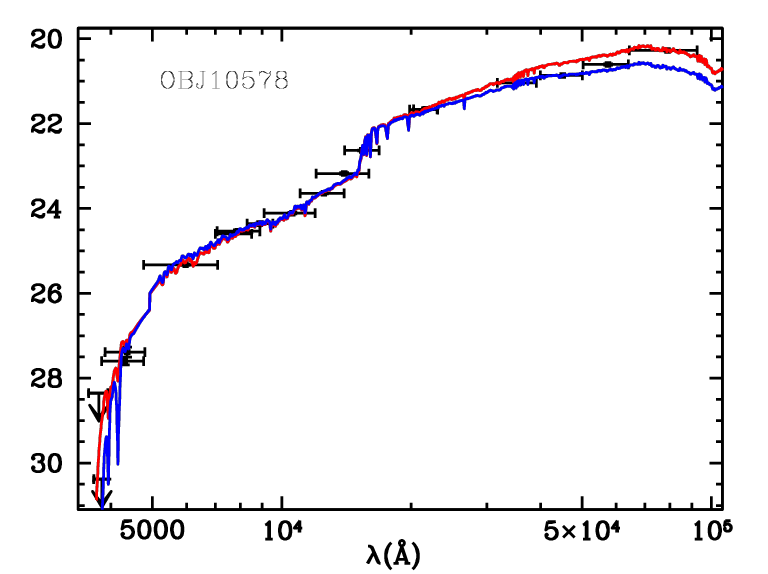}
\caption{In the top panel we show the shifts on the $UVJ$ diagram of the objects in the reference sample, when fitted with different libraries of models. Blue dots: exponentially declining ($\tau$) library; red squares: TH library. The shifts are small ($\sim$0.3 $\Delta$mag) and typically more pronounced in the $V-J$ colour. The bottom panels show two examples explaining the described trend: the fitted SEDs substantially coincide in the optical - NIR, but differ in the IRAC region. See text for more details.} \label{uvjth}
\end{figure}

In this Section we compare our results with those that can be obtained using other selection criteria (namely the rest-frame $UVJ$ diagram and the observed $VJL$ and $iHM$ diagrams), and with those obtained in similar recent studies.

\subsection{The $UVJ$ selection}

It is interesting to check which sources would be identified as passive using a more standard approach. To this aim, we started anew from the G13 catalogue, and perform the SED--fitting on the  photometric dataset, again keeping the official CANDELS redhisfts, but now using a typical library of $\tau$-models, without the inclusion of nebular lines. On this sample we tested both the standard \textit{UVJ} criterion as well as a selection based on the specific star--formation rates.

Fig. \ref{uvj} is a $UVJ$ diagram in which, for the sake of clarity, we only plot the $z_{CANDELS}>3$ sources. In this plot we use the rest-frame colours obtained from the $\tau$-model fitting. 
% Yellow points are the whole $z>3$ sample; cyan stars are quiescent objects, having sSFR $< 1/t_{U}(z)$ yr$^{-1}$ ($t_{U}(z)$ is the age of the Universe at redshift $z$). Blue dots are the TH candidates in the selection without nebular lines, and empty squares mark the ones in the selection including nebular lines emission. Finally, empty circles refer to the objects selected in the standard $\tau$-models selection. 
We display the position in the \textit{UVJ} plane of the objects in our reference sample (blue dots; the 10 objects also selected with the library including nebular emission are highlighted with empty squares). Here we use the rest-frame colours of the best-fit models; we checked that using colours computed interpolating the observed fluxes (shifted at the redshift of the source), e.g. using \textsc{EAZY}, yields qualitatively similar results, although with a larger scatter in the distribution.
Reassuringly, it is clear that most of the objects that are selected as passive with the \textit{UVJ} approach are also selected with our technique. However, some contamination is present, as a few objects fall inside this region but are not selected in our reference sample: these objects are discarded in our case due to the existence of possible star--forming solutions with $p>5\%$. 
In addition, a non negligible number of our candidates in the reference sample fall outside (below) of the passive region, in the region where recently quenched sources are expected to lie (as discussed in Sect. \ref{UVJtheor}). This confirms that the adoption of simple $\tau$-models and colour criteria may fall short in singling out a complete sample of red and dead objects, and we therefore conclude that the choice of a more sounded SFH analytic shape like the top-hat we adopted in this study can have significant impact on the selection of realistic candidates of passively evolving objects at high redshift.

In the same Fig. \ref{uvj}, we also code the objects according to their estimated sSFR in the $\tau$ fit. We selected as red and dead candidates the objects having specific rates $sSFR < 10^{-11}$ yr$^{-1}$: with this criteria, and again requiring $z_{CANDELS}>3$ and \textit{Ks}+3.6+4.5 1 $\sigma$ detection, we single out only 10 objects, marked as open circles in the Figure.
We notice that only 5 out of 10 are present in the TH selection: they are IDs 2782, 7526, 8785, 17749 and 18180. Their best-fit values are similar to those obtained with the TH libraries. 
Snapshots showing the other 5 sources included in the $\tau$ selection and \emph{not} in the reference TH selection are shown in Fig. \ref{snaps_taunotTH}; we note that ID 34275 is only clearly visible in the $H$-band image and might be a spurious detection from a close-by star, while IDs 2032, 5501, 22515 and 34636 have not been included in the TH selection despite having best fits as passive objects, because star forming solutions with $p>5\%$ are present. 

Interestingly, none of the five $z>4$ red and dead candidates in the reference selection is identified as passive with the $\tau$-models criteria. Two of them, IDs 3912 and 23626, are fitted as sources of $\sim 1.3$ Gyr and $\sim 500$ Myr respectively, missing the selection because of estimated sSFR slightly higher than the chosen threshold ($6.3 \times 10^{-11}$ and $4.2 \times 10^{-11}$ yr$^{-1}$). The other three objects (IDs 5592, 6407, 9209) are fitted as young (age $< 800$ Myr) star-forming sources with sSFR $> 10^{-10}$ yr$^{-1}$. The $\chi^2$ of the fits with the $\tau$-models and the TH models are similar. % (see the table in Appendix A). 
Again, if our modeling is correct all these are good examples of the kind of objects discussed in Section \ref{UVJtheor}: young galaxies in the early Universe which have quenched their short SF activity abruptly, just before the time they are observed, and are identified as still (slightly) star--forming in a standard $\tau$ fit because of the limitation of the chosen fitting model. 

Fig. \ref{uvjth} (top panel) displays the shifts in the $UVJ$ diagram positions for the objects belonging to the reference sample, when fitted with different libraries of models. The shifts are typically small, of the order of 0.1-0.3 $\Delta$mag (generally consistent with the uncertainties in the relevant observed colors). Noticeably, they tend to affect the $V-J$ colour more than the $U-V$ colour. This is basically due to the fact that the fit is much more robustly constrained in the region of the observed visible and NIR (covered by the ACS and WFC3 bands), which straddle the rest--frame $U-V$ break at $z\sim3$, than in the reddest part of the spectrum, since the two 5.6 to 8.0 $\mu$m IRAC bands have the poorest S/N. This allows larger variations in the $V-J$ colour, as shown by the two examples reported in the bottom panels of the same Figure. We also note a systematic effect between the two libraries, as most of the candidates with a red $[U-V]_{rest}$ have bluer $[V-J]_{rest}$ using the TH than using the $\tau$ library, while galaxies with bluer $[U-V]_{rest}$ are shifted towards redder $[V-J]_{rest}$ colours - implying a shallower mid--IR profile in all cases, as it can be seen in the two SEDs examples.
To definitively solve this ambiguity, deeper data longward of 3$\mu$m are necessary;  this anticipates the need for \textit{JWST} observations, that will be the target of the Section \ref{jwst}.

%%% Fig 1
\begin{figure}
\centering
\includegraphics[width=9cm]{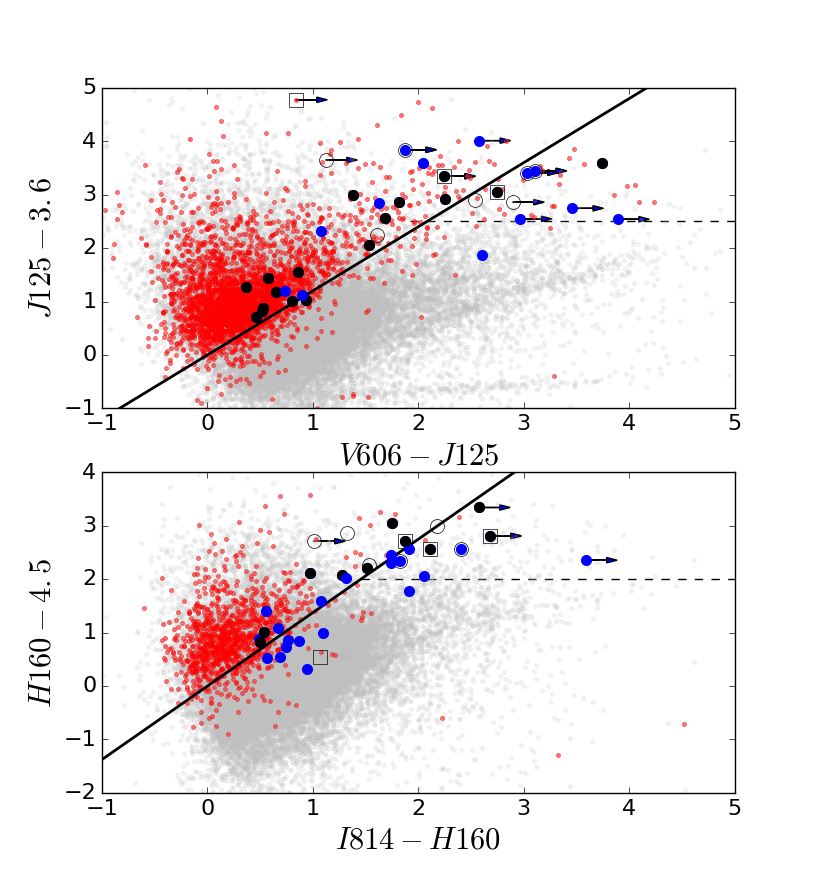}
\caption{Diagnostic colour-colour diagrams. In both panels, grey dots are the whole G13 catalogue, and red dots are the sources having $z_{CANDELS}$ in the interval of interest for the corresponding diagnostic diagram: $2.5<z<3.5$ for the $VJL$ plot (top panel), $3.5<z<4.5$ for the $iHM$ plot (bottom panel). Black and blue dots are the reference sample red and dead candidates, respectively having redshift within the interval of interest and outside it (arrows are upper-limits). Empty squares and circles refer to the $\tau$ selection (Sect. \ref{others}), again respectively having redshift within the interval of interest and outside it.} \label{colour_sel}
\end{figure}

\subsection{Diagnostic planes with observed colours} \label{diagplanes}

Fig. \ref{colour_sel} shows the result of the selections on diagnostic $VJL$ and $iHM$ observed colour-colour planes, which are the equivalent of the $BzK$ diagram \citep{Daddi2004} for selection of quiescent galaxies at $z \sim 3$ and $z \sim 4$. In each diagram, the upper right region (delimited by the diagonal solid line and the horizontal dashed line) is expected to be populated by passive sources. The grey dots are individual objects from the whole G13 catalogue, while red dots are galaxies having $z_{CANDELS}$ in the interval of interest for the corresponding diagram. Larger filled dots are the reference sample sources, again colour-coded depending on their photo-$z$ (see the caption of the Figure). While some of the selected objects lie in the passive region of the diagram, many others are found having slightly bluer observed colours. Therefore, a straightforward colour selection would exclude them from the sample \citep[see][ for similar discussions on the $BzK$ selection]{Grazian2007}.

\subsection{Comparison with previous samples}

Another interesting comparison can be made with the results from previous published studies. \citet{Rodighiero2007} used the \citet{Giavalisco2004} multi-wavelength imaging data to extract photometric data, performed a magnitude selection requiring no detection in HST bands, $K>$23.5, and IRAC 3.6 $\mu$m$<23.26$, and identified 20 objects as massive galaxies with high probability of being high-redshift, passive sources (with 14 of them also having a lower redshift, dust-obscured star forming solution). We can now check the nature of those objects using our new deeper photometry. Using the $H$-detected catalogue, we can identify 18 out of 20 sources via spatial cross-correlation.
As it turns out, none of these objects has a strong passive solution in our new analysis; the conclusion can be strengthened analyzing the SEDs of these objects, even obtained with the $\tau$ library (they generally show very weak 4000 {\AA} breaks, blue band detections and rising FIR flux), and by the cross-correlation with the 24 $\mu$m catalogue by \citet{Magnelli2013}, with 12 out of 18 objects having an association with a 24 $\mu$m prior within 0.6". Clearly the classification by \citet{Rodighiero2007} was heavily affected by the lower quality of the imaging data available at that time.

%*** FORSE PER?? VALE LA PENA DI METTERE QUALCHE IMMAGINE A SUPPORTO? ***

We then check the correspondence between our selection and two of the most recent similar works, S14 and N14. 
S14 use a $UVJ$ criterion to single out 6 quiescent candidates in the GOODS-South field, which we cross-correlate with the G13 catalogue. Among these, three (CANDELS IDs 4503, 17749, 18180) belong to our reference sample as well, while the other three (IDs 5479, 6294 and 19883) have a star--forming best fit. Indeed, they are assigned rather high sSFR in the S14 fit too  ($27.5\times10^{-11}$,  $18.6\times10^{-11}$ and $4.47\times10^{-11}$ yr$^{-1}$, respectively); they are also flagged as probable AGNs in the \citet{Cappelluti2016} catalogue, and the first two are identified as AGNs by \citet{Xue2011} as well; they all have confirmed spectroscopic redshift consistent with the $z_{CANDELS}$ we use \citep{Szokoly2004}. 

N14 identify 16 evolved (post-starburst) galaxies using a $H-K$ colour selection to probe the 4000 {\AA} break - a selection that in principle include both star--forming and passive galaxies. Five objects in their selection also belong to our reference sample (IDs 2782, 7526, 12178, 17749, 18180). Of the other eleven objects in the N14 selection, two (IDs 9177 and 16671) have $z_{CANDELS}<3$, one (ID 6189) is a low-redshift ($z_{CANDELS}=0.6$) dust-obscured star-forming galaxy in the CANDELS catalogues while it is fitted as a passive $z=4.0$ object by N14, and eight (IDs 4356, 4624, 9286, 10479, 12360, 13327, 18694 and 19195) have star--forming best fits in our analysis; five of them are also identified as AGNs in the catalogue by \citet{Cappelluti2016} \citep[with three also included in the][catalogue]{Xue2011}.

It is interesting to note that none of the previous cited works include our best candidates, IDs 10578 and 22085, in their selections. S14 only include sources at $z$$>$3.4, while ID10578 has $z_{CANDELS}$=3.06 and ID22085 has $z_{CANDELS}=$3.36. On the other hand, both galaxies fail N14 colour selection criteria ($[Y-J]$ vs. $[H-Ks]$). In the case of ID22085,  $J105$ band photometry is not available in the CANDELS GOODS-South dataset (this actually shows one more point of strength of the SED fitting approach, in that the lacking of one band data does not compromise the whole study of one potentially interesting object); in ID10578, the object falls immediately outside the selection area of their colour-colour diagram.

%Fig. \ref{ssfr_z} summarizes the result of the selections on a sSFR vs. $z$ diagram. We use the ED models fitting to plot the whole catalogue, with yellow dots. The horizontal black line gives the $z=3$ selection limit. The cyan stars are the 1001 quiescent sources, selected as described in Sect. \ref{Q}. The blue large dots are the the \textit{top-hat} passive selection, while the ED models passive selection is plotted as large void circles.

% hat said, it must be taken into account that the SED--fitting method has its own intrinsic limitations (e.g., the choice of the IMF or the stellar population synthesis models have a great impact on the final fit), but most noticeably it can suffer from the current lack of sufficiently detailed and robust photometric data. We will discuss this point in the next Section.

\section{Looking forward: the JWST perspective} \label{jwst}

\begin{figure*}
\centering
\includegraphics[width=18cm]{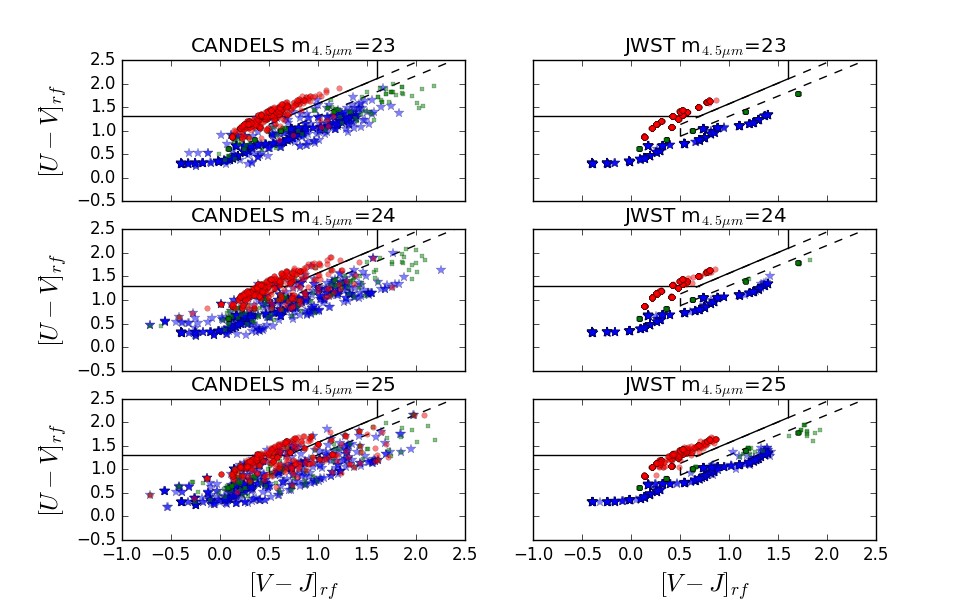}
\caption{Comparison of the $UVJ$ diagrams from mock observed catalogues, where the rest-frame magnitudes are obtained via SED--fitting using CANDELS (left panels) and \textit{JWST} (right panels) filter sets. The mock catalogues have been created starting from the TH library of spectra, simulating 1686 objects including passive and star forming galaxies, computing the observed fluxes in all the relevant bands rescaled to three reference magnitudes (each row in the Figure corresponds to one of them - from top to bottom, m$_{4.5 \mu m}$=23, 24 and 25), and including observational noise. In each panel, models having ongoing SF activity in the input library are plotted as blue stars, recently quenched objects as green squares, and passively evolving galaxies as red dots. The dashed lines define a ``green valley'' used to quantify the contamination between the different samples. It is clear that the \textit{JWST} pass-bands set removes almost completely the contamination in the fitted colours between the three different populations, which is severe in the CANDELS case. See text for more details.} \label{JWSTsim}
\end{figure*}

As our study shows, there are still many sources of uncertainty that conspire to make the search for passive objects at $z>3$ problematic and the selection uncertain: depending on the tightness of the selection criteria, one may end up with very different samples (e.g., in our case we can go from 30 to 2 objects). 
In particular, the spectral range centred on the 4000 {\AA} break is crucial, both to determine with good accuracy the photometric redshift but especially to distinguish between the star--forming and the passive objects; at $z>3$, inferring the spectral slope both below and above the 4000 {\AA} break demands a good coverage of the whole wavelength range from the $J$ to the redder \textit{Spitzer} bands, ideally up to $8\mu m$. 

The \textit{James Webb Space Telescope} appears to be perfectly suited to fill these gaps. The NIRCam and MIRI instruments\footnote{See the webpage \textit{http://www.stsci.edu/jwst/instruments/} for full information.} will include a large set of filters in the near to mid-infrared wavelength range, allowing for a detailed photometric reconstruction of the mentioned important spectral features and, hopefully, for a much easier disentanglement between degenerate solutions from SED--fitting.

It is interesting to try a rough evaluation of the potential of the \textit{JWST} capabilites in this context. To this aim, we have created a sample of synthetic spectra using our TH library. The full sample consists of 1686 simulated objects, of which 828 correspond to star--forming models (having age $< \Delta t_{burst}$), 230 have quenched the SF activity since less than 100 Myr, and 628 are red and dead (age $> \Delta t_{burst}$+100 Myr). Each of these spectra has been placed at redshifts from 3 to 7 (with the additional constraint that the age of the galaxy is not larger than the age of the Universe at that redshift) at steps of 0.1 in redshift. We then created observational catalogs corresponding to such models, reproducing both the filter sequence and depths of the CANDELS catalogue used in this work, as well as an idealized catalog reproducing a possible survey executed with \textit{JWST}. To this purpose we have replaced all the CANDELS filters redward of $Y$ (included) with a combination of twelve \textit{JWST} bands ($F090W$, $F115W$, $F150W$, $F200W$, $F277W$, $F356W$, $F444W$, $F560W$, $F770W$, $F1000W$, $F1130W$, $F1280W$), as described in the MIRI and NIRCam documentation webpages. The resulting catalog mimicks a survey executed (redward of F090) with \textit{JWST} on the GOODS-S field, building upon the existent ACS data. In particular, we created three catalogues by normalizing the magnitudes to three reference values, $m_{4.5 \mu m}$ = 23, 24 and 25, covering the magnitude range of our candidates. 
Noise has been added to these catalogs accordingly to the observed S/N versus magnitude relation  in the CANDELS filters \citep[see][]{Castellano2012}; in the \textit{JWST} simulated bands, we have assumed the depth expected in the case of an extragalactic survey for high redshift galaxies described in \citet{Finkelstein2015}. For the three reddest \textit{JWST} filters that were included there, we have computed the expected signal--to-noise assuming a total exposure time per filter comparable to each of the other \textit{JWST} filters.

These simulations are clearly simplified, since {\it a)} they use the same library to compute the ``true'' galaxy colours and to derive their photomeric redshifts and SED properties from the SED--fitting, and {\it b)} because we ignore on the one hand the additional gain to the overall photometry that will be possible using the improved resolution of \textit{JWST} compared to \textit{Spitzer}, and on the other hand any possible complication due to blending of sources and other systematics. Regardless of these limitations, these tests can give us a preview of the improvements that \textit{JWST} will make possible.

We have repeated on these simulated catalogs the same analysis that we did on real data. 
We first fitted catalogues with our SED--fitting code, and then computed the rest--frame properties at the photometric redshift. For simplicity, we show here the results obtained in the $UVJ$ plane, for the CANDELS-like and the \textit{JWST}-like catalogs separately. They are shown in Fig. \ref{JWSTsim}, for all objects having 3<$z_{phot}$<7. In each panel, models having ongoing SF activity in the input library are plotted as blue stars, recently quenched objects as green squares, and passively evolving galaxies as red dots. The results on the CANDELS simulated data show that there is a strong contamination in the fit, as many passive and star--forming galaxies end up in the same regions of the $UVJ$ diagram. For example, one can define a ``green valley'' as the region of the diagram for which $V-J>0.5$ and $0.88\times[V-J]+0.44<U-V<0.88\times[V-J]+0.69$ (see the dashed lines in Fig. \ref{JWSTsim}): considering the three datasets with reference magnitudes $m_{4.5 \mu m}$ = 23, 24 and 25, the CANDELS simulation respectively yields $\sim$3.6, 7.2 and 10.0\% star forming galaxies erroneously falling within or above the green valley; conversely, the passive models falling within or below the green valley in the three cases are $\sim$3.0, 16.7 and 31.4\%. %On the other hand, the \textit{JWST} simulations yields $\sim$0 misplaced models in all of the three cases.

This contamination increases (as expected) when input galaxies are fainter. This simulations confirm that the identification of passive galaxies in the CANDELS data set is potentially prone to misidentification due to the still inadequate depth of the photometry. 

Conversely, the situation is much more defined using the \textit{JWST} filters: the three populations are robustly fitted and separated, with almost no contamination even down to the faintest magnitudes (the observed ``arched'' distributions on the diagram derive from the input true colours, which the fitted ones closely resemble). This is an exciting demonstration of the future capabilities with the new instrument. 

It is interesting to note that some red objects again fall outside the passive region of the diagram, as discussed in the previous Sections: this shows again how the $UVJ$ colour selection can be prone to the risk of missing objects that have quenched their SF activity in recent times, even using a much more accurate photometric dataset.

\section{Summary and conclusions}\label{summary}

In this paper we have presented the methods and results of a study aimed at searching passive galaxies in the early Universe. The summary of the work is the following.

\begin{itemize}
\item We search for high--redshift, red and dead (i.e. passively evolving) galaxies in the GOODS-South field, using an updated version of the \citet{Guo2013} photometric catalog that includes CANDELS HST fluxes, HUGS $Ks$ data, and new IRAC images and improved photometric measurements (Sect. \ref{dataset}). We pre-select $H$--detected objects having \textit{Ks}, IRAC 3.6 and 4.5 $\mu$m $\geq$1 $\sigma$ detection, and $z_{CANDELS}>$3. We also add a new sample of 178 $K$/IRAC-detected sources from Boutsia et al. (in preparation) and \citet{Wang2016}.

\item We then analyze this selection using dedicated \textit{top-hat} libraries for SED--fitting. We assume that a single star formation event took place and abruptly stopped in the past, followed by passive evolution ever since, and we fit the observed fluxes with models having different values for the duration of the burst, the UV extinction, and the metallicity.  The selection criterion is based on two stringent requirements: the selected objects must have at least one passive model solution (i.e., SFR=0 and age larger than the burst duration) with $p(\chi^2)>30\%$, and do not have any star--forming solution with a probability $p(\chi^2)>5\%$. 

\item We first use a library without nebular lines emission and only consider the CANDELS redshifts. This way we select 30 candidates, all of which are $H$--detected (see Figs. \ref{snaps_TH} and \ref{SEDsTH}). 

\item Including nebular lines in the top-hat library used for the SED--fitting procedure, only 10 of these candidates survive the probabilistic selection process: in many cases, the lines weaken the fitted continuum redward of the 4000 {\AA} break, yielding a star--forming best--fit; in other cases, the probabilistic approach causes the exclusion of galaxies with alternative solutions.

\item If we repeat the analysis letting the redshift free to vary around the best-fit value,  only two galaxies (IDs 10758 and 22085) retain their passive status as the only robust solution. All the other objects show alternative star--forming solutions (at different redshifts)  with a probability $p(\chi^2)>5\%$.  

%Many objects in the TH selection have SEDs which can hardly be considered as passive at a first glance (3897, 4587, 5592, 6407, 7310, 8446, 9209, 16506, 19505); however, they are compatible with recently quenched galaxies. We find 5 passive sources at $z>4$: ID 7310 ($z>4.948$, $M=2.152 \times 10^{10} M_{\odot}$), ID 6407 ($z=4.744$, $M=1.984 \times 10^{10} M_{\odot}$), ID 23626 ($z=4.637$, $M_*=7.597 \times 10^{10} M_{\odot}$), ID 9209 ($z= 4.549$,  $M_*=6.408 \times 10^{10}$ $M_{\odot}$), ID 5592 ($z=4.448$, $M=2.559 \times 10^{10} M_{\odot}$). These objects have metallicities $Z/Z_{\odot} \simeq 0.4$ and ages between 200 and 400 Myr. Interestingly, none of these sources are identified as passive in the ED models selection.

%The three objects in the ED models selections which are \emph{not} in \textit{top-hat} selection are IDs 2032, 2869 and 34024 (Fig.\ref{snaps_taunotTH}). They can be considered as spurious detections, as discussed in Sect.\ref{others}.

\item Since it is not possible to completely rule out strongly obscured star--forming solutions for any of the selected sources (see Fig. \ref{probebvTHall}), as a basic sanity check we perform a cross-correlation of the reference sample with the 24 $\mu$m catalogue by \citet{Magnelli2013}, on \textit{Herschel} PEP-GOODS \citep{Lutz2011} and HerMES \citep{Smith2012} blind catalogues, and on \citet{Wang2016} new catalogue based on $H$-detected priors. Two objects in our selection are associated to strong FIR emitters. Interestingly, they are also identified as optical counterparts of X--ray emitters \citep{Cappelluti2016,Xue2011}; we therefore speculate that they might be recently quenched galaxies, hosting a dust-obscured AGN. No other object in the reference sample has a clear association with a FIR source.

\item By means of a direct selection on the full G13 catalogue using a standard exponential $\tau$-models fit with BC03, we then identify, for comparison, 10 sources as $z$>3 passive candidates (we require $sSFR < 10^{-11}$ yr$^{-1}$). 5 objects are in common between this selection and the reference sample (IDs 2782, 7526, 8785, 17749, 18180). %Among these, ID 8785 is likely to be a $z=3.98$ passive source (the Universe being $\sim$ 1.6 Gyr old), with a mass $3.894 \times 10^{10}$ $M_{\odot}$ and an age of 910 Myrs.

\end{itemize} 

A clear outcome of our analysis is that the selection of passive galaxies, at least in the considered range of redshifts, is still prone to significant uncertainties, due to the limitations in the assumptions used in the  SED fitting models and the relatively modest $S/N$ of the objects. Nevertheless, considering the weakest among our selection criteria we can at least derive an upper limit for the number density of these objects, finding $\sim$0.173 arcmin$^{-2}$ (or $\sim$$2.0\times10^{-5}$ Mpc$^{-3}$ for 3<$z$<5).

The limitations in the SED modeling hampers our chances to derive robust physical information on the selected sample. Ages are poorly constrained, and thus so are the SF rates necessary to assemble such objects. 
We can try some educated guess on the minimum sSFR of the selected sources (assuming isolated evolution, i.e. no mergers) by taking their estimated stellar masses, and dividing them by the age of the Universe at the time the SF activity ceased (minus 300 Myr, to crudely exclude the dark ages), in the TH best--fit models (we consider the fit without nebular lines, for simplicity). This yields a typical lower threshold for the sSFR of $\sim 7 \times 10^{-10}$ yr$^{-1}$, which is fairly consistent with the observed values of main-sequence star forming galaxies, in the same redshift and mass regimes \citep[e.g.][]{Salmon2015, Schreiber2017}. This is the sSFR estimated \emph{at the end} of the activity, i.e. when the mass has been completely assembled; we note that, since in our scheme the rate of star formation of the models is constant before the quenching, if we had observed the galaxies during the star formation phase they would have been classified as starbursts, because having lower stellar mass they would be \emph{above} the Main Sequence $\dot{M}-M$ relation.
%, a factor of $\sim ...$ higher than the estimated sSFR of coeval main sequence galaxies \citep{...}.

% Our probabilistic selection method, based on the analysis of the SED--fitting process in which models with \textit{top-hat} (abrutply quenched) SFHs were used, allowed for the identification of 25 more objects with respect to a standard $\tau$-models selection, at the same time excluding some spurious identifications; furthermore, it has ensured a more robust and complete selection of reliable candidates with respect to standard colour-based methods. 
% On the other hand, the adoption of different and increasingly demanding criteria reduces the final sample to 10 candidates (including emission lines in the fit), or to just two sources (leaving $z$ free to vary in the fit), showing how major uncertainties still affect this kind of search.

%4 sources have been excluded from our TH selection despite having a passive best fit, because of the presence of star forming solution with non-negligible probability.

By means of a dedicated simulation, we have shown how \textit{JWST} will yield a major improvement in this perspective, allowing for a much more effective detachment of high-$z$ passive objects from dust-obscured low-$z$ ones, thanks to an effective coverage of crucial regions of the observed spectra - namely, the 4000 {\AA} break and the 20 $\mu$m rest-frame regions.

A thorough testing against theoretical expectations for the number density and properties of these kind of objects, at the considered redshifts, is compelling and recommended.

%%%% Fig 2
%\begin{figure*}[h] 
%\centering
%\includegraphics[width=10cm]{figs/colour_sel_1.png}
%\caption{colour-colour diagrams using Guo photometry to identify passive candidates. In all panels, yellow dots are the whole sample, green dots are sources with IRAC 3.6 magnitude $<$ 23.15. Red triangles show the positions of Rodighiero's sources using their original photometry; they move to the black stars positions using Guo's photometry. Blue circles are the $\tau$-models selection and open circles are the \textit{top-hat} candidates.} \label{colour_sel1}
%\end{figure*}

%%%% Fig 10
%\begin{figure*}[h] 
%\centering
%\includegraphics[width=14cm]{figs/sSFRebv1.png}
%\includegraphics[width=14cm]{figs/sSFRebv2.png}
%\caption{Specific SFR vs. UV extinction of the models for 4 objects. IDs 2717, 2782 and 7526 belong to both FITEXP and FITTH. ID 2032 is only in FITEXP and shows large evidence of possible ongoing SF activity. }\label{sSFRebv}
%\end{figure*}

%%% Figs 12 13 14
%\begin{figure*}[h] 
%\centering
%\includegraphics[width=17cm]{figs/Rodighiero1.png}
%\caption{SED--fitting ($\tau$-models) for the Rodighiero objects identified in the Guo catalogue.}\label{rodsed1}
%\end{figure*}
%\begin{figure*}[h] 
%\centering
%\includegraphics[width=17cm]{figs/Rodighiero2.png}
%\caption{SED--fitting ($\tau$-models) for the Rodighiero objects  identified in the Guo catalogue.}\label{rodsed2}
%\end{figure*}

\section*{Acknowledgements}
The research leading to these results has received funding from the European Union
Seventh Framework Programme (FP7/2007-2013) under grant agreement n. 312725. We thank the anonymous referee for the useful comments and suggestions, which greatly helped to improve the paper.

%%%%%%%%%%%%%%%%%%%%%%%%%%%%%%%%%%%%%%%%%%%%%%%%%%



\bibliographystyle{mnras}
\bibliography{biblio} % if your bibtex file is called example.bib


%%%%%%%%%%%%%%%%%%%%%%%%%%%%%%%%%%%%%%%%%%%%%%%%%%

%%%%%%%%%%%%%%%%% APPENDICES %%%%%%%%%%%%%%%%%%%%%
\onecolumn
\appendix

%\begin{landscape}
\section{Physical properties of the selected sample of red and dead candidates}

\begin{longtable}{ | l || l | l | l | l |}
\hline
ID$_{CANDELS}$ & $z_{CANDELS}$  & $\chi_{reduced}^2$ & Age [Gyr] & Stellar mass [$10^9$ M$_{\odot}$] \\ \hline\hline
10578 & 3.06 & 1.97 & $0.63_{-0.31}^{+0.37}$ & $239.70_{-54.40}^{+108.80}$  \\ \hline
22085 & 3.36 & 1.26 & $0.61_{-0.30}^{+0.97}$ & $44.25_{-12.31}^{+17.08}$  \\ \hline
2717 & 3.04 & 1.13 & $1.58_{-0.97}^{+0.00}$ & $162.70_{-69.01}^{+56.80}$  \\ \hline
2782 & 3.47 & 0.94 & $0.71_{-0.40}^{+0.88}$ & $69.95_{-26.15}^{+16.27}$  \\ \hline
3912 & 4.08 & 1.32 & $1.25_{-0.94}^{+0.05}$ & $36.27_{-15.79}^{+25.08}$  \\ \hline
8785 & 3.98 & 0.72 & $0.91_{-0.59}^{+0.39}$ & $38.96_{-15.65}^{+17.36}$  \\ \hline
9209 & 4.55 & 1.61 & $0.41_{-0.21}^{+0.74}$ & $91.51_{-44.79}^{+27.69}$  \\ \hline
17749 & 3.73 & 0.63 & $0.90_{-0.39}^{+0.40}$ & $108.80_{-51.19}^{+30.20}$  \\ \hline
18180 & 3.61 & 1.36 & $0.91_{-0.50}^{+0.39}$ & $90.11_{-37.97}^{+22.29}$  \\ \hline
23626 & 4.64 & 1.05 & $0.41_{-0.11}^{+0.69}$ & $75.91_{-25.90}^{+28.89}$  \\ \hline
2608 & 3.58 & 1.36 & $0.63_{-0.43}^{+0.42}$ & $4.51_{-1.83}^{+1.23}$  \\ \hline
3897 & 3.14 & 1.12 & $0.36_{-0.16}^{+0.69}$ & $11.86_{-1.82}^{+10.14}$  \\ \hline
3973 & 3.67 & 1.84 & $0.91_{-0.30}^{+0.39}$ & $186.10_{-85.20}^{+16.40}$  \\ \hline
4503 & 3.52 & 3.74 & $1.10_{-0.80}^{+0.20}$ & $142.70_{-59.61}^{+38.20}$  \\ \hline
4587 & 3.58 & 2.55 & $0.41_{-0.21}^{+0.85}$ & $5.48_{-1.72}^{+4.34}$  \\ \hline
5592 & 4.45 & 1.05 & $0.36_{-0.16}^{+0.79}$ & $30.16_{-16.22}^{+17.90}$  \\ \hline
6407 & 4.74 & 1.31 & $0.36_{-0.16}^{+0.69}$ & $15.98_{-3.71}^{+11.26}$  \\ \hline
7526 & 3.42 & 0.60 & $0.90_{-0.58}^{+0.68}$ & $36.24_{-17.45}^{+17.89}$  \\ \hline
7688 & 3.35 & 0.69 & $0.61_{-0.31}^{+0.97}$ & $22.76_{-11.71}^{+12.09}$  \\ \hline
8242 & 3.18 & 1.17 & $1.00_{-0.69}^{+0.58}$ & $6.55_{-2.25}^{+1.97}$  \\ \hline
9091 & 3.30 & 2.22 & $0.36_{-0.16}^{+0.90}$ & $2.81_{-0.81}^{+2.69}$  \\ \hline
10759 & 3.07 & 1.38 & $0.63_{-0.62}^{+0.95}$ & $0.91_{-0.64}^{+1.17}$  \\ \hline
12178 & 3.28 & 1.02 & $1.10_{-0.79}^{+0.20}$ & $41.20_{-10.46}^{+16.64}$  \\ \hline
15457 & 3.41 & 1.98 & $0.36_{-0.16}^{+0.69}$ & $4.39_{-0.63}^{+2.90}$  \\ \hline
16506 & 3.34 & 3.68 & $0.36_{-0.16}^{+0.69}$ & $5.06_{-0.67}^{+3.62}$  \\ \hline
19301 & 3.60 & 2.85 & $1.10_{-0.90}^{+0.20}$ & $11.58_{-5.39}^{+7.36}$  \\ \hline
19446 & 3.25 & 2.89 & $1.00_{-0.80}^{+0.58}$ & $20.07_{-10.73}^{+3.37}$  \\ \hline
19505 & 3.33 & 1.20 & $0.63_{-0.43}^{+0.57}$ & $46.63_{-15.00}^{+6.19}$  \\ \hline
22610 & 3.22 & 0.74 & $0.63_{-0.43}^{+0.63}$ & $9.49_{-3.09}^{+4.40}$  \\ \hline
26802 & 3.45 & 1.75 & $0.63_{-0.43}^{+0.95}$ & $4.67_{-1.92}^{+2.45}$  \\ \hline
\caption{Physical properties of the red and dead candidates belonging to the reference sample, as obtained from their best fit with the TH library without emission lines. ID$_{CANDELS}$ is the identification number in the G13 catalogue; $z_{CANDELS}$ is the official CANDELS redshift. $\chi_{reduced}^2$ is the normalized (reduced) $\chi^2$ of the best fit. The SFR is always zero, by definition.
The table lists first the two most robust candidates, which have passes all the selection criteria including the free $z$ fit; second, the other 8 objects, identified as passive in the emission line fit as well (see table \ref{tab2}); finally, the remaining 20 objects in the reference sample.}\label{tab1}
\end{longtable}

\newpage

\begin{longtable}{ | l || l | l | l | l |}
\hline
ID$_{CANDELS}$ & $z_{CANDELS}$  & $\chi_{reduced}^2$ & Age [Gyr] & Stellar mass [$10^9$ M$_{\odot}$] \\ \hline\hline
10578 & 3.06 & 1.97 & $0.63_{-0.47}^{+0.37}$ & $239.60_{-83.80}^{+108.70}$  \\ \hline
22085 & 3.36 & 1.26 & $0.61_{-0.59}^{+0.97}$ & $44.23_{-33.61}^{+17.04}$  \\ \hline
2717 & 3.04 & 1.13 & $1.58_{-0.97}^{+0.00}$ & $162.50_{-62.89}^{+56.70}$  \\ \hline
2782 & 3.47 & 0.94 & $0.71_{-0.40}^{+0.88}$ & $69.90_{-26.12}^{+16.26}$  \\ \hline
3912 & 4.08 & 1.32 & $1.25_{-0.94}^{+0.05}$ & $36.26_{-15.78}^{+25.07}$  \\ \hline
8785 & 3.98 & 0.72 & $0.91_{-0.59}^{+0.39}$ & $38.95_{-15.65}^{+17.35}$  \\ \hline
9209 & 4.55 & 1.61 & $0.41_{-0.40}^{+0.74}$ & $91.49_{-77.82}^{+27.61}$  \\ \hline
17749 & 3.73 & 0.63 & $0.90_{-0.39}^{+0.40}$ & $108.80_{-51.29}^{+30.10}$  \\ \hline
18180 & 3.61 & 1.36 & $0.91_{-0.50}^{+0.39}$ & $90.04_{-38.04}^{+22.26}$  \\ \hline
23626 & 4.64 & 1.05 & $0.41_{-0.11}^{+0.69}$ & $75.88_{-25.89}^{+28.92}$  \\ \hline
\caption{Physical properties of the 10 red and dead candidates passing the probabilistic selection including nebular lines emission in the fit. ID$_{CANDELS}$ is the identification number in the G13 catalogue; $z_{CANDELS}$ is the official CANDELS redshift. $\chi_{reduced}^2$ is the normalized (reduced) $\chi^2$ of the best fit. The SFR is always zero, by definition. The table lists first the two most robust candidates, which have passes all the selection criteria including the free $z$ fit; then, the other 8 objects identified as passive in the emission line fit.}\label{tab2}
\end{longtable}

\section{Snapshots of the TH candidates}

%\begin{figure*}[h!] 
\centering
\includegraphics[width=15.5cm]{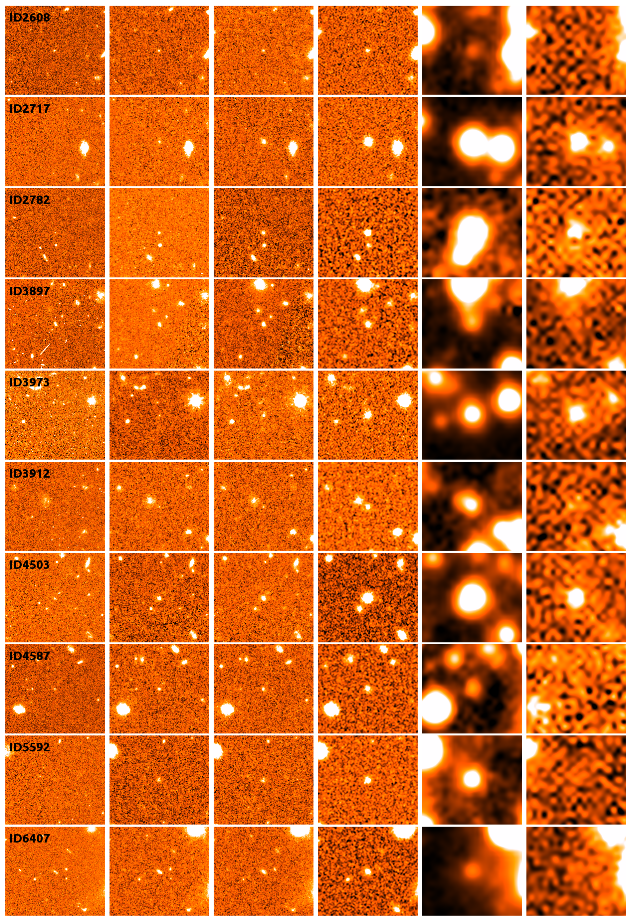}
\newpage
\includegraphics[width=15.5cm]{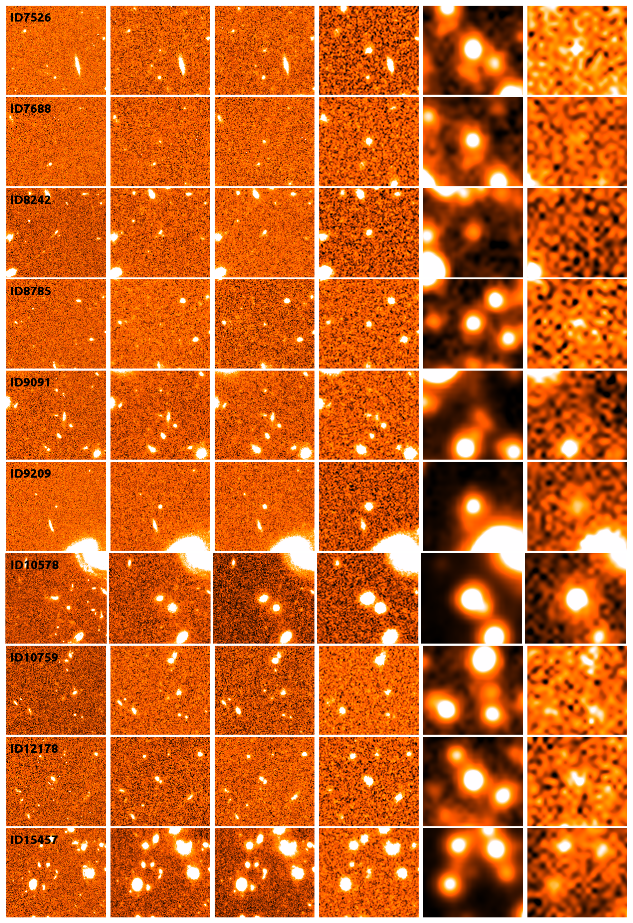}
\newpage
\includegraphics[width=15.5cm]{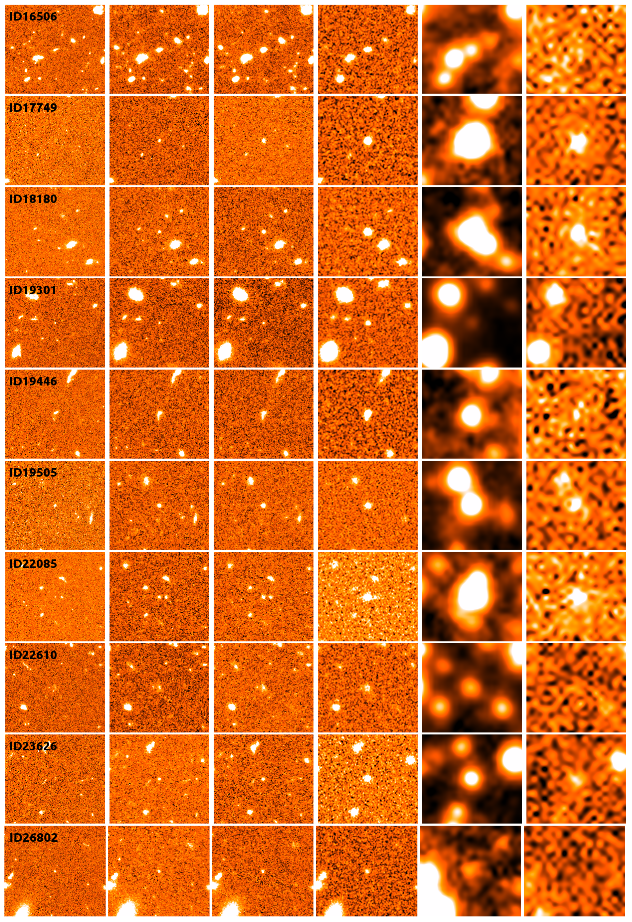}
\captionof{figure}{Snapshots of the 30 passive candidates selected in the reference sample, obtained with the TH library. Left to right: $ACS$ $ B435+V606+I814$ stack, $WFC3$ $J125$, $WFC3$ $H160$, Hawk-I $Ks$, IRAC $3.6 + 4.5$ $\mu$m stack, IRAC $5.8 + 8.0$ $\mu$m stack.}\label{snaps_TH}
%\end{figure*}

\newpage
\section{SEDs of the TH reference sample candidates}

%%% Fig 8a

\centering
\includegraphics[width=7cm]{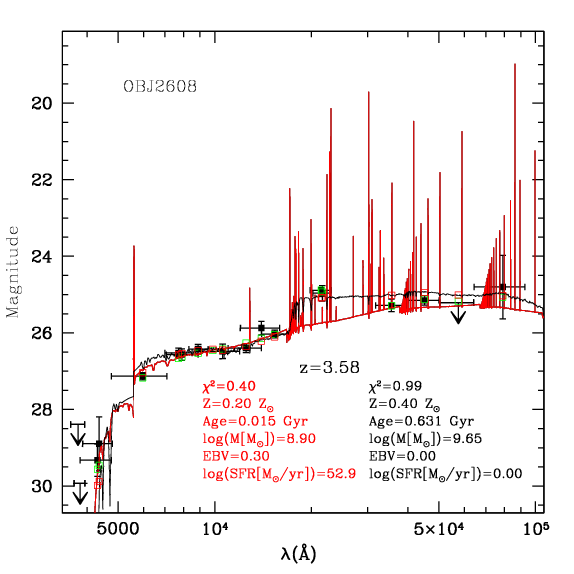}
\includegraphics[width=7cm]{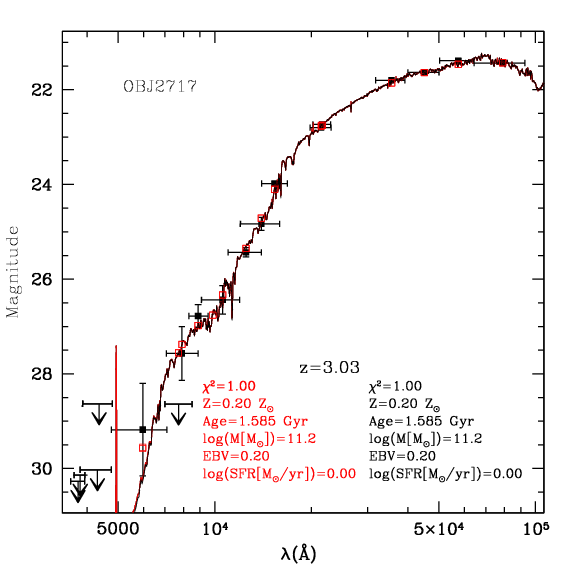}
\includegraphics[width=7cm]{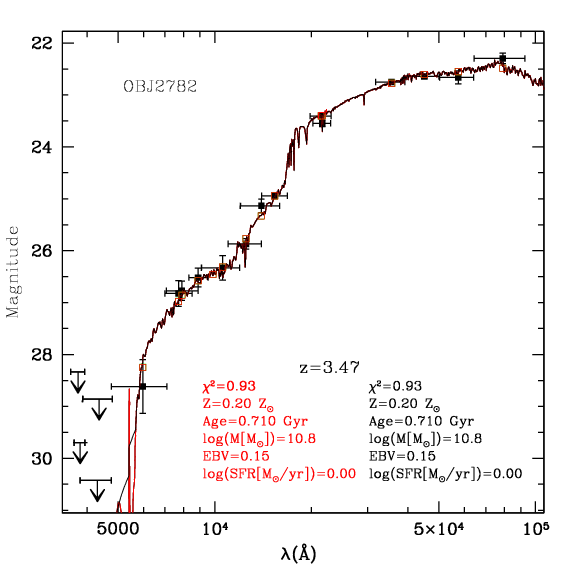}
\includegraphics[width=7cm]{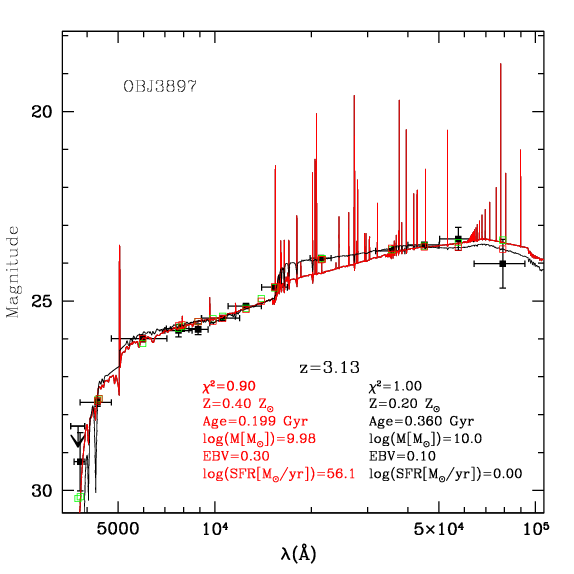}
\includegraphics[width=7cm]{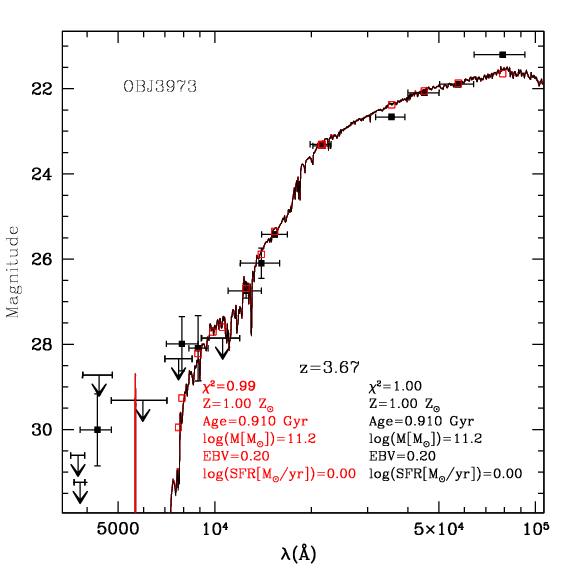}
\includegraphics[width=7cm]{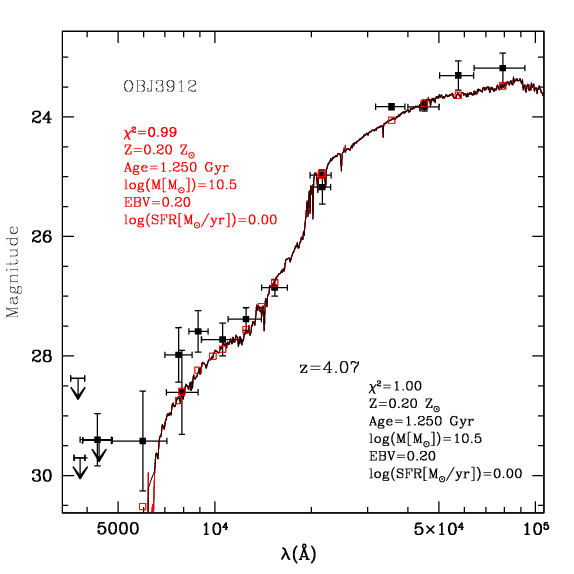}
\includegraphics[width=7cm]{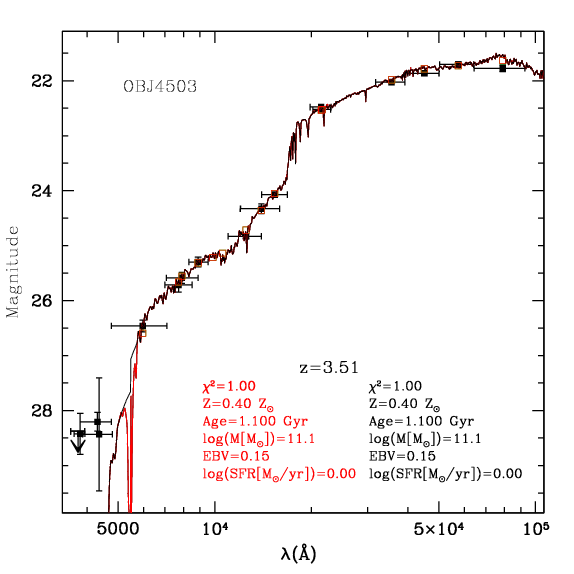}
\includegraphics[width=7cm]{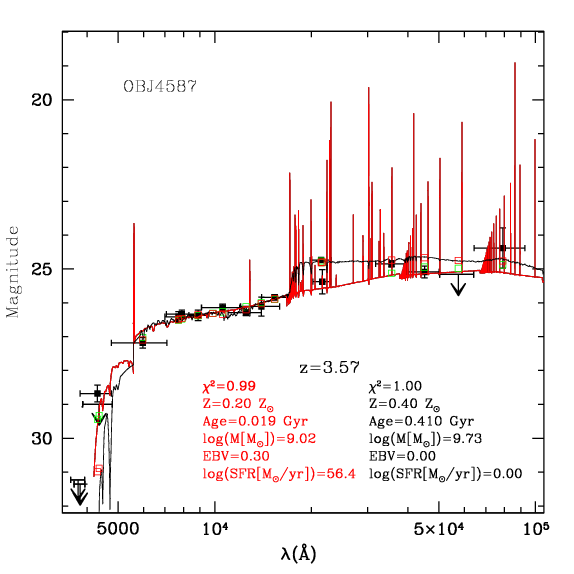}
\includegraphics[width=7cm]{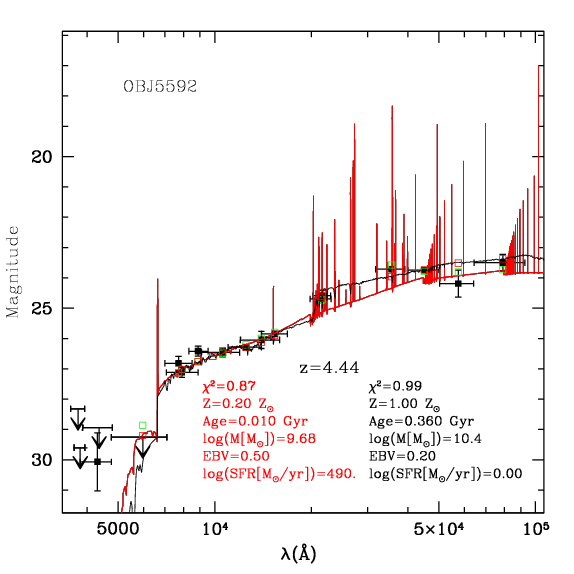}
\includegraphics[width=7cm]{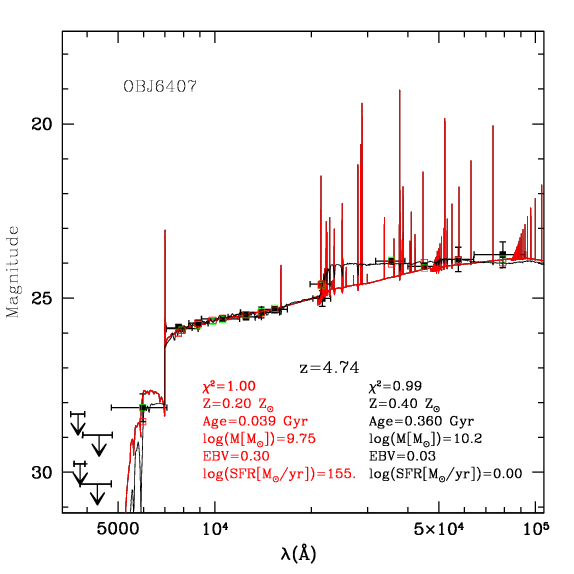}
\includegraphics[width=7cm]{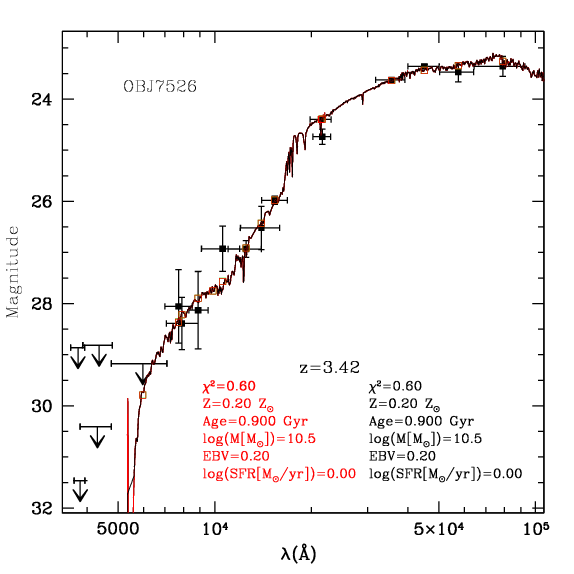}
\includegraphics[width=7cm]{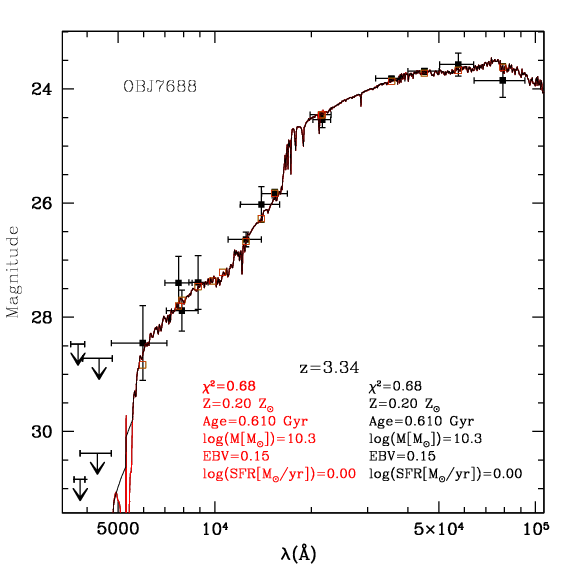}
\includegraphics[width=7cm]{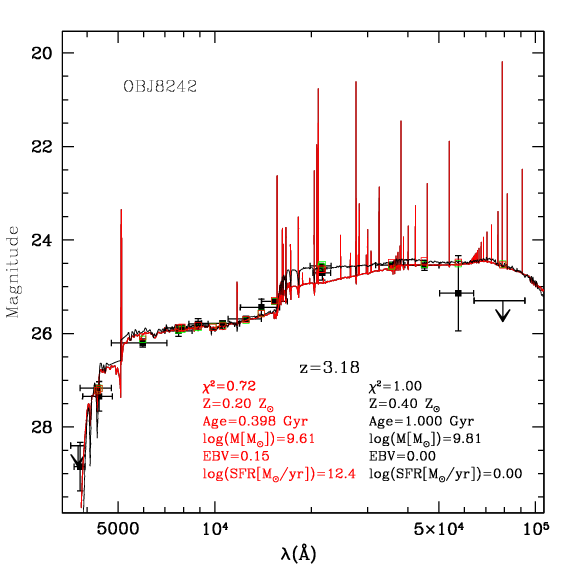}
\includegraphics[width=7cm]{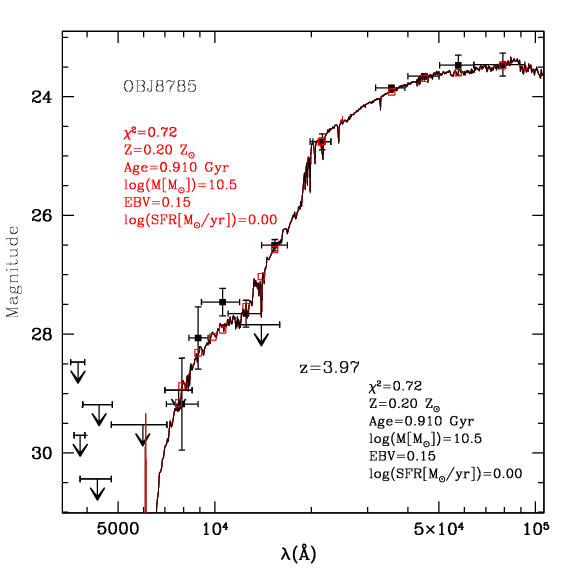}
\includegraphics[width=7cm]{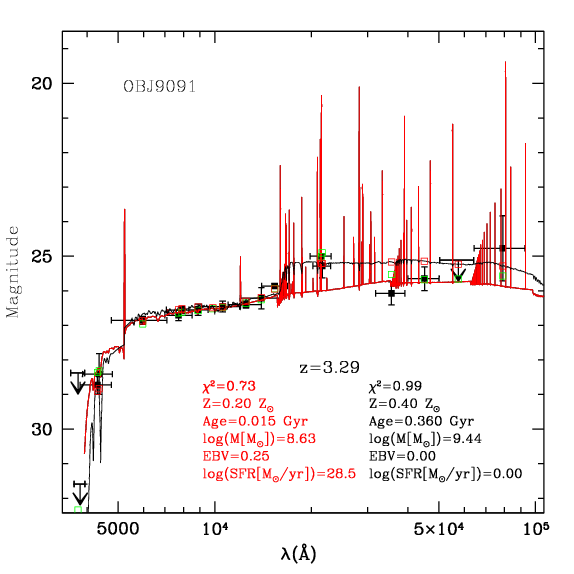}
\includegraphics[width=7cm]{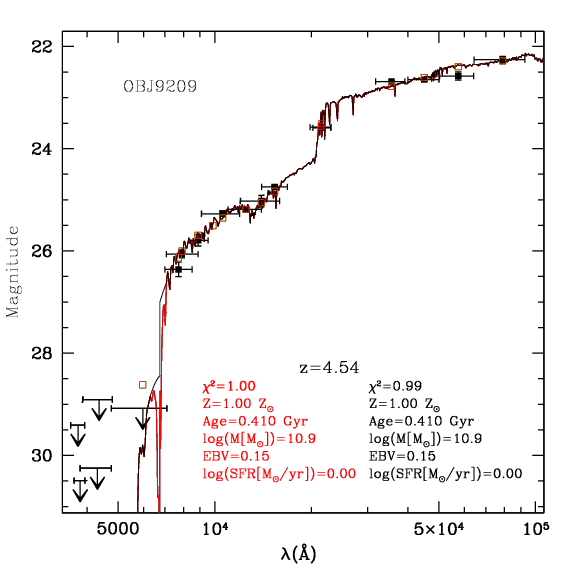}
\includegraphics[width=7cm]{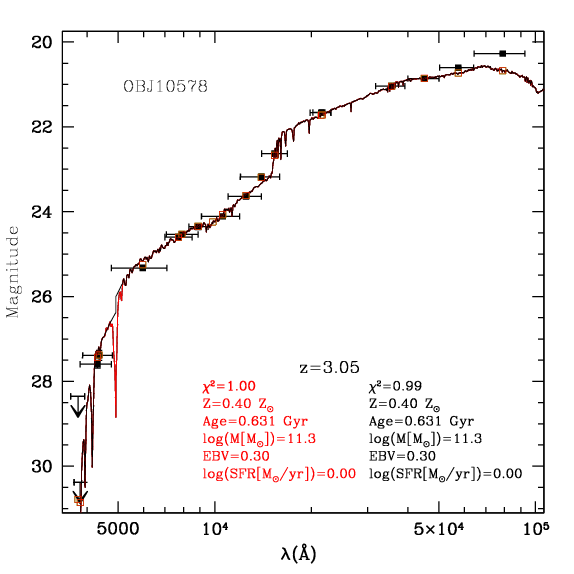}
\includegraphics[width=7cm]{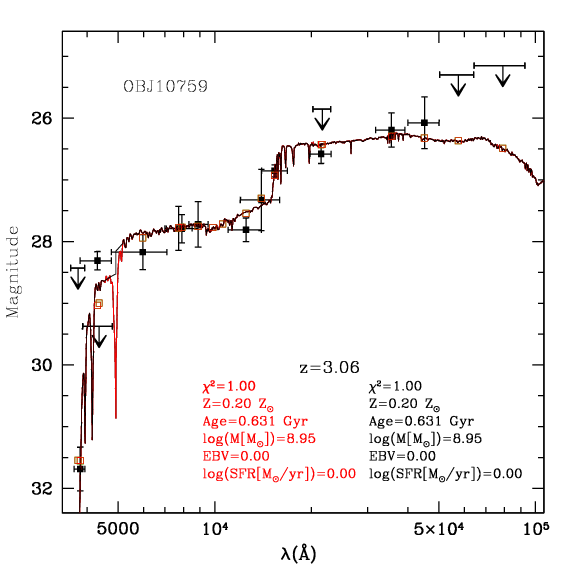}
\includegraphics[width=7cm]{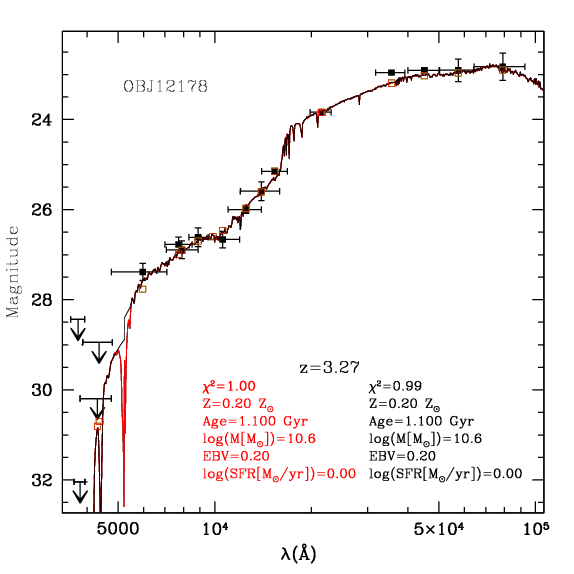}
\includegraphics[width=7cm]{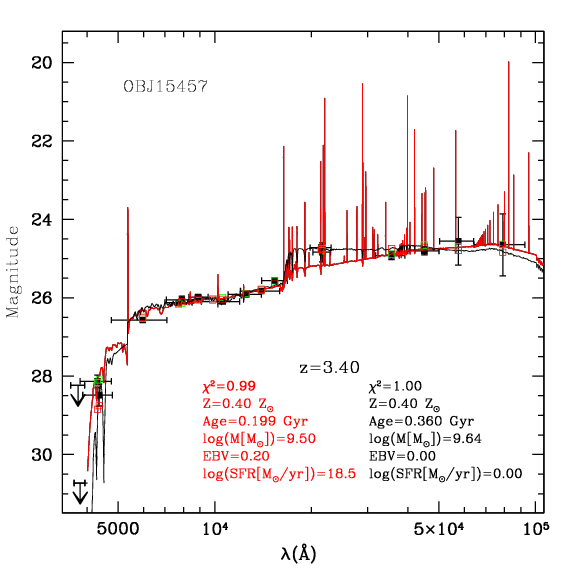}
\includegraphics[width=7cm]{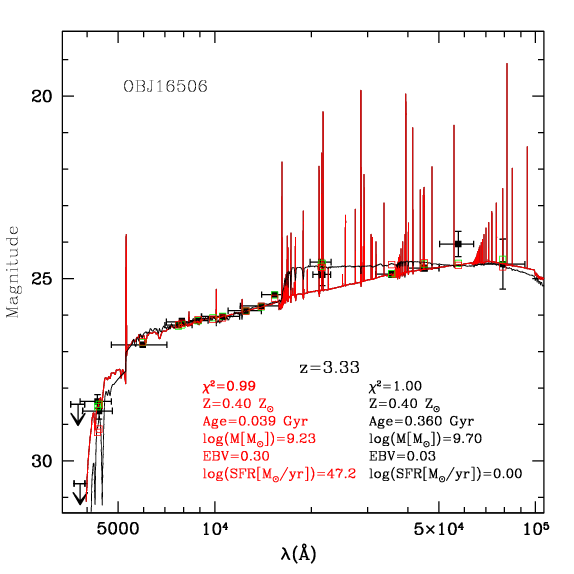}
\includegraphics[width=7cm]{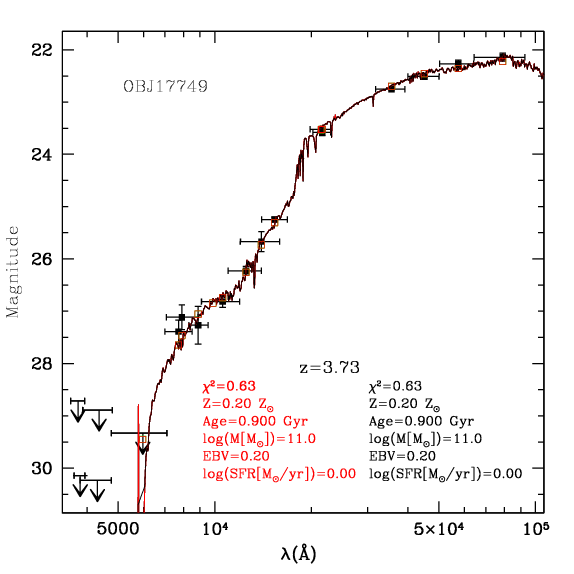}
\includegraphics[width=7cm]{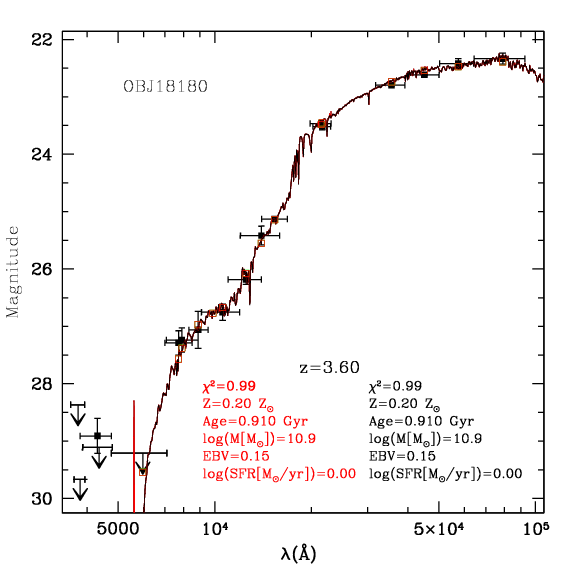}
\includegraphics[width=7cm]{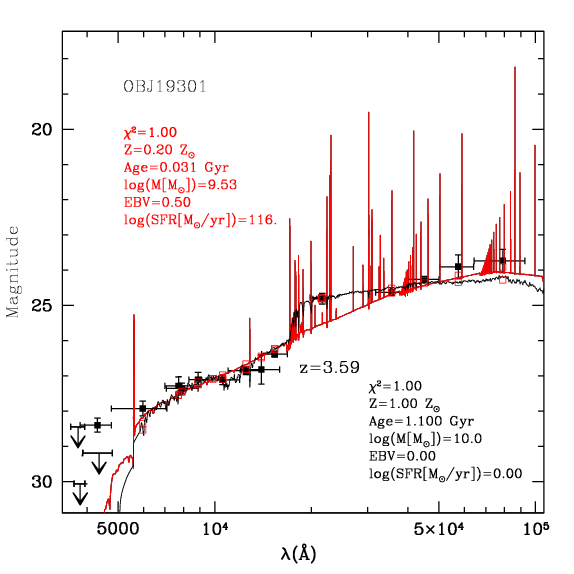}
\includegraphics[width=7cm]{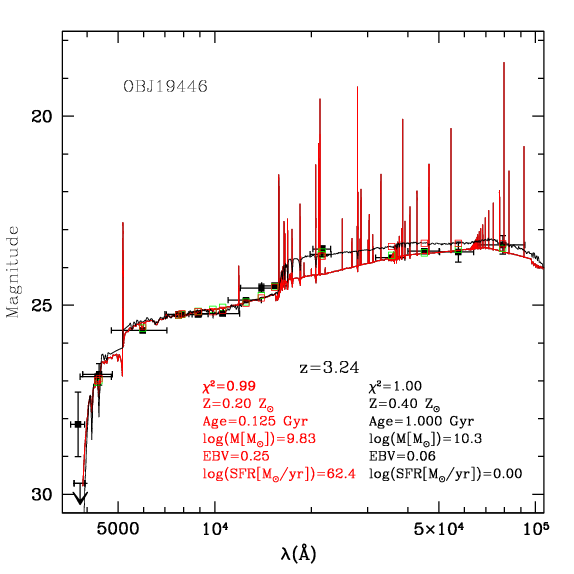}
\includegraphics[width=7cm]{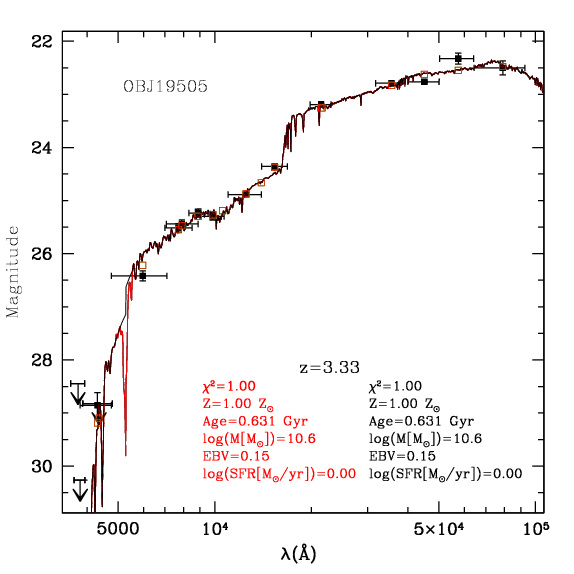}
\includegraphics[width=7cm]{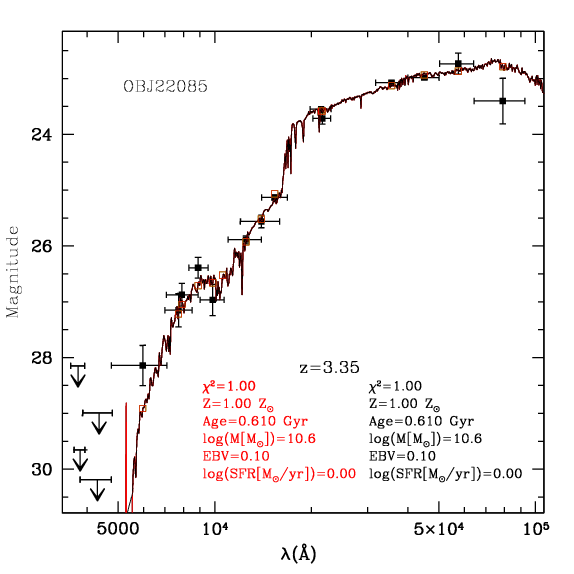}
\includegraphics[width=7cm]{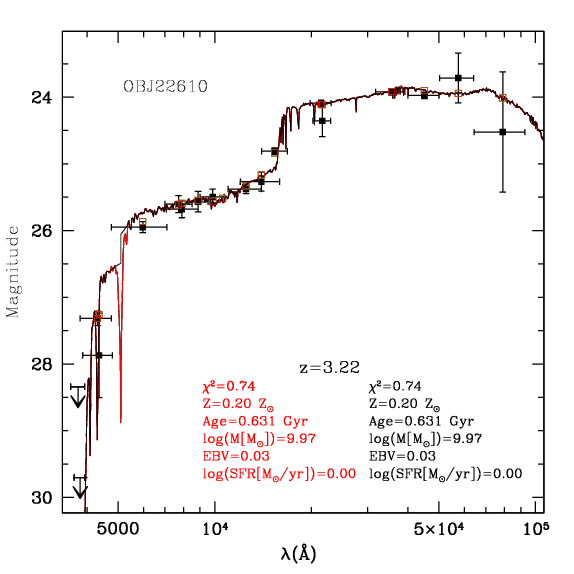}
\includegraphics[width=7cm]{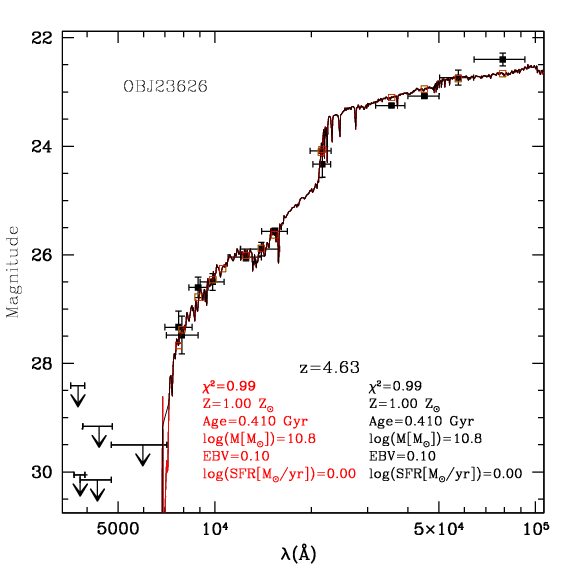}
\includegraphics[width=7cm]{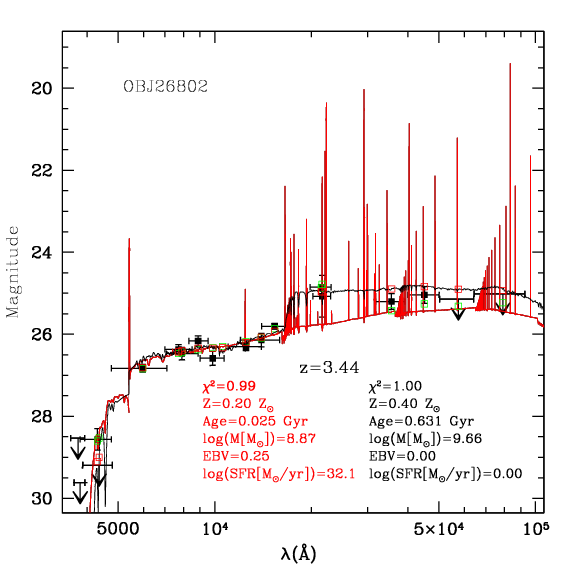}

\captionof{figure}{SED--fitting for the objects in the reference sample. Shown is the best fit using the TH libraries, with $z=z_{CANDELS}$, with (red line) and without (black line) the inclusion of nebular emission; in many cases the two fits almost coincide, so the two lines are superposed. The physical parameters of the best fit models are reported on the bottom of each plot, with colours (blue or black) corresponding to the considered fit.} \label{SEDsTH}
%\end{figure*}

\section{Probability and extinction of all the model solutions of the TH candidates} \label{probs}
\centering
\includegraphics[width=14cm]{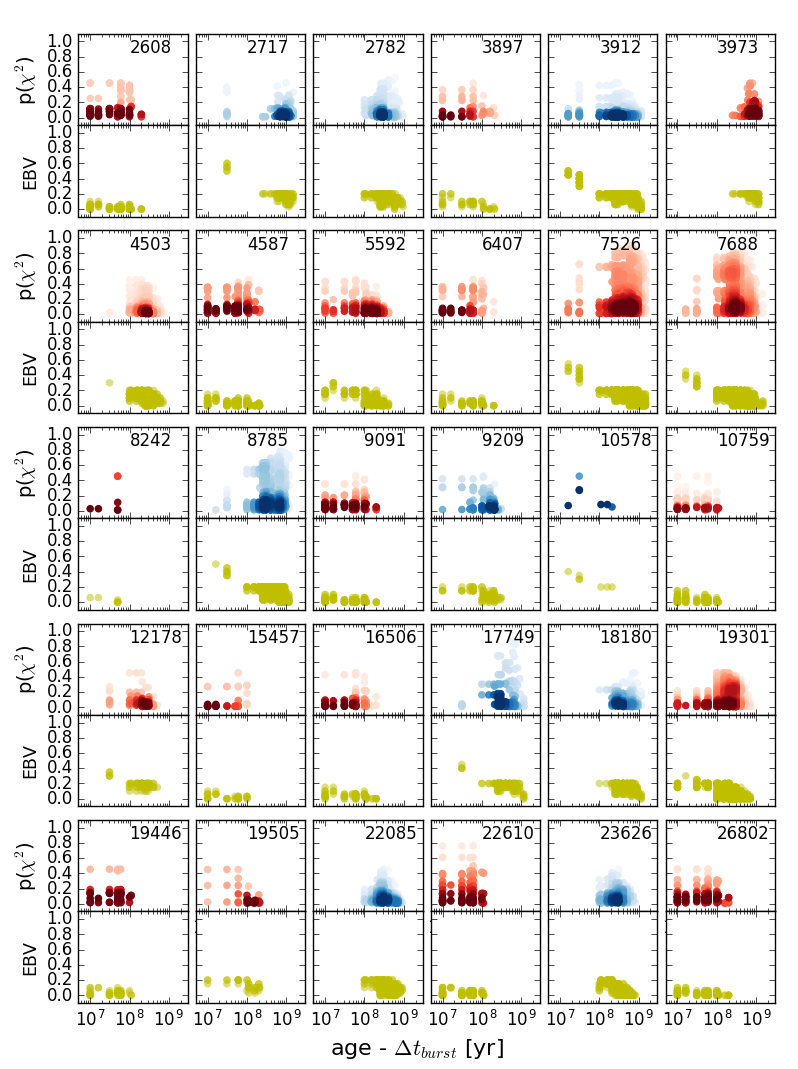}
\captionof{figure}{Probability and dust extinction as a function of [age - $\Delta t_{burst}$] for all the possible solutions in the SED--fitting process, for all the candidates in the reference sample. For each candidate, indicated by its ID, two panels are shown. In the upper one, dots represent the probability $p(\chi^2)$ of each model solution; the colours of the dots refer to the belonging of the source to the selection with (blue) or without (red) the inclusion of nebular lines; the dots (i.e. the models) are shaded as a function of their density. The lower one shows the corresponding values of the $E(B-V)$. All the solutions have age $> \Delta t_{burst}$, as required to be classified as passive in this approach. Galaxies excluded from the selection, on the other hand, have been fitted by at least one model with age $\leq \Delta t_{burst}$, i.e. still star--forming, with a probability $p>5\%$ (not shown).}\label{probebvTHall}

\newpage
\section{Snapshots of the $\tau$-models candidates}
%\begin{figure*}[h!] 
\centering
\includegraphics[width=15cm]{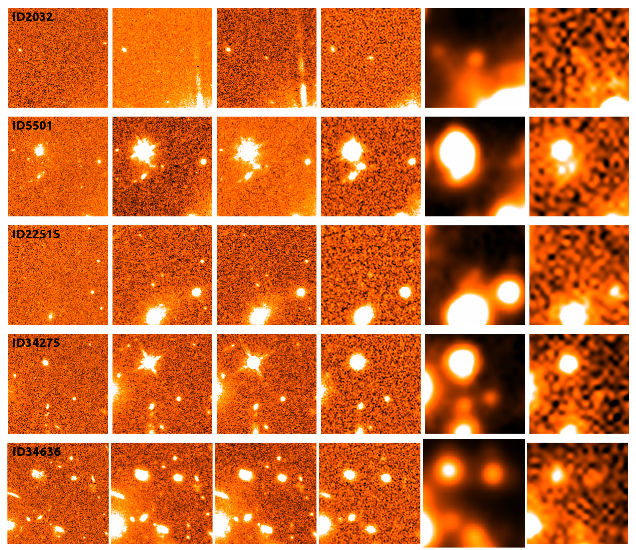}
\captionof{figure}{Snapshots of the 5 passive candidates selected with the $\tau$-models library which are not present in the reference sample. Left to right: $ACS$ $ B435+V606+I814$ stack, $WFC3$ $J125$, $WFC3$ $H160$, Hawk-I $Ks$, IRAC $3.6 + 4.5$ $\mu$m stack, IRAC $5.8 + 8.0$ $\mu$m stack.}\label{snaps_taunotTH}
%\end{figure*}

%%%%%%%%%%%%%%%%%%%%%%%%%%%%%%%%%%%%%%%%%%%%%%%%%%

% Don't change these lines
\bsp	% typesetting comment
\label{lastpage}
\end{document}